\documentclass[]{aa}
\usepackage{graphicx}
\usepackage{txfonts}
\newcommand{\kms}{\,km\,s$^{-1}$}

\newcommand{\dpcvn}{DP\,CVn}
\newcommand{\dipsc}{DI\,Psc}
\newcommand{\Halpha}{H$\alpha $ }

\newcommand{\vsini}{{$v\sin i$}}

\begin{document}

   \title{Doppler imaging of stellar surface structure}

   \subtitle{XXIV. The lithium-rich single K-giants  \dpcvn\ and \dipsc
   \thanks{Based on observations obtained at the Kitt Peak
   National Observatory, U.S.A., and at the Canada-France-Hawaii Telescope,
   U.S.A.}}

\author{Zs.~K\H{o}v\'ari\inst{1}
  \and H.~Korhonen\inst{2,3}
  \and K.~G.~Strassmeier\inst{4} 
  \and M.~Weber\inst{4} 
  \and L.~Kriskovics\inst{1}
  \and I.~Savanov\inst{5}}

\offprints{Zs. K\H{o}v\'ari, \email{kovari@konkoly.hu}}

\institute{Konkoly Observatory of the Hungarian Academy of Sciences, Konkoly Thege \'ut 15-17., H-1121, Budapest, Hungary\\
  \email{kovari@konkoly.hu}
  \and Niels Bohr Institute, University of Copenhagen, Juliane Maries Vej 30, DK-2100 K{\o}benhavn \O , Denmark
  \and Centre for Star and Planet Formation, Natural History Museum of Denmark, University of Copenhagen, {\O}ster Voldgade 5-7, DK-1350 K{\o}benhavn \O , Denmark
  \and Leibniz Institute for Astrophysics (AIP), An der Sternwarte 16, D-14482 Potsdam, Germany
  \and Institute of Astronomy of the Russian Academy of Sciences, Pyatnitskaya 48, 119017, Moscow, Russia
             }

   \date{Received ; accepted}


\abstract  
   {}
{We present the first Doppler imaging study of the two rapidly rotating, single K-giants \dpcvn\ and \dipsc\ in order to study the surface
spot configuration and to pinpoint their stellar evolutionary status.}
{Optical spectroscopy and photometry are used to determine the fundamental astrophysical properties.
Doppler imaging is applied to recover the surface temperature distribution for both stars, while photometric light-curve inversions are carried out for studying the long-term
changes of the surface activity of \dpcvn. Surface differential rotation of \dpcvn\  is estimated from cross-correlating the available subsequent Doppler reconstructions separated
by roughly one rotation period.}
{Both stars appear to have higher than normal lithium abundance, LTE $\log n$ of 2.28 (\dpcvn) and 2.20 (\dipsc), and are
supposed to be located at the end of the first Li dredge-up on the RGB. Photometric observations reveal rotational modulation with a period of 14.010\,d (\dpcvn) and 18.066\,d (\dipsc). Doppler reconstructions from the available mapping lines well agree in the revealed spot patterns,
recovering rather low latitude spots
for both stars with temperature contrasts of $\Delta T\approx600-800$\,K below the unspotted photospheric background.
Spots at higher latitudes are also found but either with less contrast (\dpcvn) or
with smaller extent (\dipsc).
A preliminary antisolar-type differential rotation with $\alpha=-0.035$ is found for \dpcvn\ from cross-correlating the subsequent Doppler images.
Long-term photometric analysis supports the existence of active longitudes, as well as the differential rotation.}
{}

\authorrunning{K\H{o}v\'ari et al.}
\titlerunning{Doppler imaging of \dpcvn\ and \dipsc}

   \maketitle

%

\section{Introduction}

The surface lithium abundance is a crucial diagnostic mark for studying stellar structure and evolution. The $^{7}$Li isotope, which
represents the vast majority of the solar Li, is destroyed through proton capture at the temperatures of $\sim$3\,MK (Lugaro \& Chieffi \cite{lugaro11}).
This means that the stellar surface layers are progressively depleted of lithium through material transportation to the deeper,
hotter layers by convective motions, turbulent diffusion and meridional flows. Therefore, the surface lithium abundance can in
principle be used as an age indicator but, in reality, the situation is more complex.

Comparison of Li lines observed from the quiet Sun and sunspot umbrae shows that the Li lines are enhanced in sunspots (e.g.,
Grevesse \cite{grev}; Giampapa \cite{giam}). This is due to the lower temperature of the spots. Giampapa (\cite{giam}) suggested
that this effect could possibly also be seen in disk-integrated stellar spectra and that the strength of the lithium
line in stars could be greatly affected by cool spots. On the other hand, the solar lithium abundance is lower in hot plage
regions. This effect could compensate the effect of the apparent Li enhancement from cool spots, when looking at disk integrated spectra.
Several studies of the relation of Li abundance with stellar activity were published. These studies had different
results though, from a clear correlation (e.g., Jeffries et al. \cite{jef}, Fern{\'a}ndez \& Miranda \cite{fer}, Neuh{\"a}user et
al. \cite{neu}) to no relation at all, even in the presence of very strong starspot activity (e.g., Pallavacini et al. \cite{pal},
Mallik \cite{mallik}, Berdyugina et al. \cite{ber2}).

In a statistical analysis De~Medeiros et al. (\cite{demed}) tested the possible connection between rotation and Li abundance in single
F, G and K-giants. Their result showed a link between the rotation-rate discontinuity at spectral type G0III and the Li
abundance near the same spectral type. Giants cooler than G0III are generally very slow rotators (a mean of the \vsini\ value of 2~\kms)
and generally also show low Li abundances, while the giants warmer than G0 are fast rotators and have higher Li abundances.
Just recently, Strassmeier et al. (\cite{rotact}) found a trend of increased Li abundance with rotation period but with a dispersion of
up to 3--4 orders of magnitude. In this study, stars in binaries appear to exhibit 0.25~dex less surface lithium than singles,
as one would expect if the depletion mechanism is rotation dependent. On the
other hand, there is a small group of single K-giants which show rapid rotation and also larger than expected Li abundances (e.g., Reddy \&
Lambert \cite{red:lam2}, Drake et al. \cite{drake}, Fekel \& Balachandran \cite{fek:bal},  and references therein). Charbonnel \&
Balachandran (\cite{cha:bal}) concluded that relatively high Li abundance is not abnormal for a low-mass star on the red-giant
branch (RGB). They identified two distinct evolutionary episodes on the RGB in which extra mixing of Li may cause an unusual high
surface abundance. Both episodes are related to the close vicinity of the bottom of the convective envelope to the hydrogen-burning
shell. Rotational mixing will likely play an important role in bringing Li-rich material into the convective mantle (cf. Lugaro \& Chieffi \cite{lugaro11}).
Per definition, we call a red-giant star ``Li-rich'' once it has evolved through the first dredge-up and still has a logarithmic surface
abundance greater than 1.5.

Attention to  \dpcvn\ (HD\,109703) and \dipsc\ (HD\,217352)  was drawn in the
Vienna-KPNO Doppler-imaging candidate survey (Strassmeier et al. \cite{kgs2}). 
According to this database, the stars are rapidly rotating single K-giants, having projected rotational
velocities of $\approx$35\kms\  (with 2-4\kms\ estimated errors).
Charbonnel \& Balachandran (\cite{cha:bal}) pointed out that these two stars are new candidates for the
small group of rapidly-rotating Li-rich giants that appear at or very near the RGB bump and already exhibit freshly synthesized lithium. 
Preliminary determinations of the Li abundances yielded $\log n$(Li) of 2.82 and 2.64
for \dpcvn\ and \dipsc, respectively (Strassmeier et al. \cite{kgs2}).

In this paper, we redetermine the absolute dimensions (see Sect.~\ref{sect_param})
and the lithium abundance of these two giant stars (Sect.~\ref{sect_lithium}). We show them to have indeed a
constant radial velocity within our precision limits, thus are 
single stars (Sect.~\ref{sect_obs}). In Sect.~\ref{sect_di}, we study the surface spot configuration and
evolution by means of Doppler imaging and complement this in Sect.~\ref{sect_longphot} with a long-term photometric spot analysis.
Finally, Sect.~\ref{sect_concl} concludes on our results. 

\section{Observations}

\subsection{Spectroscopy and radial velocities}\label{sect_obs}
Most spectroscopic observations were obtained at Kitt Peak National Observatory (KPNO) with the 0.9-m coud{\'e} feed
telescope between March 29 and May 9, 2000 (summarized in Table~\ref{obs_spectra}).  The F3KB 3k-CCD was used together with
grating A, the long collimator and a 280\,$\mu$m slit. This configuration allows a resolving power of $R\approx$28,000 at a
dispersion of 4.81\kms/pixel. The full wavelength coverage was 300\,{\AA} and the spectra were centered at 6490\,{\AA}. The
exposure time was typically 4500 seconds for \dpcvn\ and 1800 seconds for \dipsc, giving average signal-to-noise ratios
(S/N) of 130:1 and 160:1, respectively.

Six high resolution spectra were obtained with the 3.6m Canada-France-Hawaii telescope (CFHT) during May 18 and 19, 2000.
The Gecko spectrograph was used with the 316 l/mm grating in 8th order and provided a resolution of $R$$\approx$120,000.
The 4.4k$\times$2k 13.5\,$\mu$m-pixel EEV1E CCD gives a $\approx$10\,nm wide wavelength range centered at the Li-6708 line.

Data reductions were done using the NOAO/IRAF software package and followed our standard procedure for coud\'e spectra (see, e.g.,
Weber \& Strassmeier \cite{web:str}). Nightly observations of various radial-velocity standards were used to obtain the radial
velocities for  \dpcvn\ and  \dipsc\  (Table~\ref{obs_spectra}). For  \dpcvn\ we used mostly 16\,Vir (K0.5III, $v_{\rm r}$=+36.48\kms) as
the standard except for the spectra at HJDs 2,451,632.65 and 2,451,633.77 when $\beta$\,Gem (K0III, $v_{\rm r}$=+3.23\kms)
was used and HJDs 2,451,663.75 and 2,451,670.84 when HD\,145001 (G8III, $v_{\rm r}$=--10.33\kms) was used.
For \dipsc\ we used mostly $\beta$\,Oph (K2III, $v_{\rm r}$=--12.18\kms) except for HJD 2,451,672.96 where we used HD\,145001,
and for the three CFHT spectra where we used $\gamma$\,Ser (F6IV, $v_{\rm r}$=+6.58\kms) and $\beta$\,CVn (G0V,
$v_{\rm r}$=+6.33\kms) (velocity standards  are taken from Scarfe et al. \cite{scarfe}). There appears to be a zero-point shift of
2\,\kms\ of the \dipsc\ velocities between KPNO and CFHT data, which we interpret to be mostly due to the use of $\beta$\,Oph
as the KPNO template star. Fekel (2004, private communication) obtained five more velocities of \dipsc\ with the coud\'e feed telescope at
KPNO and got an average velocity of --18.1\kms, in agreement with our CFHT velocities. We therefore assumed the CFHT velocities to be
correct and accordingly shifted the KPNO data by 2.0\kms. The average radial velocity values of \dpcvn\  from a
total of 23 measurements is --29.6$\pm$0.5 (rms)\kms, while from 18 measurements the average value for \dipsc\  is --18.3$\pm$0.7 (rms)\kms.
The small rms values suggest that both stars are single.

A summary of the spectroscopic observations is given in Table~\ref{obs_spectra}.

\begin{table*}
\caption{Summary of the spectroscopic observations for \dpcvn\ and \dipsc. The table gives the HJDs of the
observations, the corresponding phases calculated using Eq.~\ref{eqdpcvn} and Eq.~\ref{eqdipsc} from Sect.~\ref{sect_rot}, respectively,
and the measured signal-to-noise (S/N) ratio of the spectra, the wavelength region ($\Delta\lambda$), the heliocentric radial velocity values
with their errors
and the observatory.}\label{obs_spectra}
\centering
\begin{tabular}{c c c c c c | c c c c c c}
\hline
\multicolumn{6}{c}{\dpcvn} & \multicolumn{6}{|c}{\dipsc}\\
\hline
HJD & phase & S/N & $\Delta\lambda$& $v_{\rm helio}$ & observing & HJD & phase & S/N & $\Delta\lambda$& $v_{\rm helio}$ & observing \\
2,451,000+ &  &   &  [{\AA}]  & [km/s] & site & 2,451,000+ &  &   &   [{\AA}]  & [km/s]  &  site   \\
\hline
238.92 & 0.90 & - & 3895--4110 & - & KPNO & 068.82 & 0.44 & - & 3895--4109 & - & KPNO \\
240.85 & 0.04 & - & 6470--6787 & - & KPNO & 070.78 & 0.55 & - & 6469--6787 & - & KPNO \\
632.63 & 0.00 & 145 & 6331--6650 & -29.3$\pm$0.4 & KPNO & 639.00 & 0.00 & 133 & 6331--6650 & - & KPNO \\
633.75 & 0.08 & 100 & 6331--6650 & -29.6$\pm$0.5 & KPNO & 642.01 & 0.17 &   89 & 6331--6650 & -18.6$\pm$0.5 & KPNO \\
636.66 & 0.29 & 154 & 6331--6650 & -30.4$\pm$0.4 & KPNO & 661.00 & 0.22 & 160 & 6331--6650 & -16.5$\pm$0.4 & KPNO \\
637.75 & 0.37 & 148 & 6331--6650 & -29.3$\pm$0.6 & KPNO & 661.99 & 0.27 & 166 & 6331--6650 & -18.5$\pm$0.6 & KPNO \\
638.70 & 0.43 & 150 & 6331--6650 & -29.7$\pm$0.5 & KPNO & 662.99 & 0.33 & 164 & 6331--6650 & -18.0$\pm$0.6 & KPNO \\
639.79 & 0.51 & 147 & 6331--6650 & - & KPNO & 663.99 & 0.38 & 152 & 6331--6650 &-18.1$\pm$0.5 & KPNO \\
640.72 & 0.58 & 130 & 6331--6650 & -29.3$\pm$0.4 & KPNO & 665.99 & 0.49 & 167 & 6331--6650 & -17.2$\pm$0.4 & KPNO \\
641.70 & 0.65 & 138 & 6331--6650 & -29.3$\pm$0.4 & KPNO & 666.99 & 0.55 & 161 & 6331--6650 & -18.2$\pm$0.4 & KPNO \\
642.77 & 0.72 & 131 & 6331--6650 & -29.1$\pm$0.5 & KPNO & 667.99 & 0.60 & 156 & 6331--6650 & -18.4$\pm$0.6 & KPNO \\
644.72 & 0.86 & 122 & 6331--6650 & -28.4$\pm$0.7 & KPNO & 668.99 & 0.66 & 190 & 6331--6650 & -18.4$\pm$0.7 & KPNO \\
660.79 & 0.01 & 134 & 6331--6650 & -29.9$\pm$0.5 & KPNO & 669.98 & 0.71 & 177 & 6331--6650 & -18.7$\pm$0.6 & KPNO \\
661.77 & 0.08 & 138 & 6331--6650 & -29.7$\pm$0.5 & KPNO & 670.99 & 0.77 & 149 & 6331--6650 & -17.5$\pm$0.7 & KPNO \\
662.77 & 0.15 & 129 & 6331--6650 & -29.8$\pm$0.4 & KPNO & 671.96 & 0.82 & 168 & 6331--6650 & -18.1$\pm$0.5 & KPNO \\
663.76 & 0.22 & 132 & 6331--6650 & -29.6$\pm$0.5 & KPNO & 672.97 & 0.88 & 106 & 6331--6650 & -19.1$\pm$0.7 & KPNO \\
664.80 & 0.30 & 123 & 6331--6650 & -29.7$\pm$0.4 & KPNO & 673.94 & 0.93 & 152 & 6460--6779 & -18.5$\pm$0.6 & KPNO \\
665.78 & 0.37 & 112 & 6331--6650 & -29.3$\pm$0.6 & KPNO & 673.99 & 0.94 & 206 & 6331--6650 & -18.5$\pm$0.5 & KPNO \\
666.78 & 0.44 & 114 & 6331--6650 & -29.0$\pm$0.5 & KPNO & 683.12 & 0.44 & 230 & 6385--6460 & - & CFHT \\
667.79 & 0.51 & 118 & 6331--6650 & -30.2$\pm$0.5 & KPNO & 683.13 & 0.44 & 230 & 6385--6460 & - & CFHT \\
668.79 & 0.58 & 139 & 6331--6650 & -29.6$\pm$0.4 & KPNO & 683.13 & 0.44 & 230 & 6385--6460 & - & CHFT \\
669.78 & 0.65 & 126 & 6331--6650 & -29.7$\pm$0.4 & KPNO & 684.12 & 0.50 & 250 & 6654--6760 & -19.5$\pm$1.7 & CFHT\\
670.85 & 0.73 & 135 & 6331--6650 & -29.4$\pm$0.5 & KPNO & 684.13 & 0.50 & 250 & 6654--6760 & -18.8$\pm$2.1 & CFHT\\
671.75 & 0.79 & 103 & 6331--6650 & -29.8$\pm$0.5 & KPNO & 684.14 & 0.50 & 250 & 6654--6760 & -18.5$\pm$2.2 & CFHT \\
672.71 & 0.86 & 131 & 6331--6650 & -30.7$\pm$0.9 & KPNO & & & & & \\ 
673.75 & 0.94 & 116 & 6331--6650 & -30.6$\pm$1.0 & KPNO  \\
\hline
\end{tabular}
\end{table*}

\subsection{Photometry}
Photometric observations of \dpcvn\ and \dipsc\ were obtained with Wolfgang and Amadeus, the two 0.75-m twin automatic
photoelectric telescopes (APTs) of the Leibniz-Institute for Astrophysics Potsdam (formerly owned by the University of Vienna)
at Fairborn observatory in Arizona (Strassmeier et al. \cite{kgs1}).
Amadeus is optimized for red wavelengths and used Johnson-Cousins $VI_{C}$ filters while Wolfgang is optimized for blue
wavelengths and used Str\"omgren $by$ filters. All the measurements were done differentially in respect to HD\,109529 for
\dpcvn\ and to HD\,217019 for \dipsc. Check stars were HD\,108400 and HD\,217590, respectively. For further details on the
observing procedure and data reduction we refer to Strassmeier et~al. (\cite{kgs1}) and Granzer et~al. (\cite{woozi2}). The
observations of \dpcvn\ consist of a total of 1169 data points, with the largest number of data points (381) in the $V$ band. For
\dipsc\ we have in total 346 measurements with the largest number (128) in the $y$ band. Mean photometric errors by bandpasses
are 0$\fm$006 ($V$), 0$\fm$007 ($I_{\rm C}$),  0$\fm$004 ($b$), and 0$\fm$003 ($y$) for both stars in this paper. A summary of the photometric
observations is given in Table~\ref{obs_photo}.

\begin{table*}
\caption{Photometric observation of \dpcvn\ and \dipsc\ obtained
with the Wolfgang and Amadeus APTs. The table gives the HJDs of the observations with the corresponding UT dates
and the total number of observations in each bandpass. }\label{obs_photo}
\centering
\begin{tabular}{c c c c c c | c c c c }
\hline
 & & \multicolumn{4}{c|}{\dpcvn} & \multicolumn{4}{c}{\dipsc} \\
\hline
HJD & UT date & $V$ & $I_{\rm C}$ & $b$ & $y$ & $V$ & $I_{\rm C}$ & $b$ & $y$ \\
\hline
2451098.7--2451197.6 & 12.10.1998--19.01.1999 & 0 & 0 & 0 & 0 & 0 & 0 & 0 & 89 \\
2451250.0--2451360.7 & 12.03.1999--01.07.1999 & 0 & 0 & 0 & 68 & 0 & 0 & 0 & 0 \\
2451434.8--2451539.6 & 13.09.1999--27.12.1999 & 0 & 0 & 0 & 0 & 74 & 73 & 0 & 0 \\
2451551.9--2451825.8 & 08.01.2000--08.10.2000 & 104 & 97 & 134 & 133 & 16 & 14 & 38 & 39 \\
2451879.0--2452079.8 & 30.11.2000--14.06.2001 & 44 & 40 & 45 & 47 & 0 & 0 & 0 & 0 \\
2452177.8--2452179.8 & 25.09.2001--27.09.2001 & 0 & 0 & 0 & 0 & 2 & 1 & 0 & 0 \\
2452248.0--2452423.7 & 04.12.2001--29.05.2002 & 80 & 81 & 0 & 0 & 0 & 0 & 0 & 0 \\
2452629.0--2452786.7 & 20.12.2002--27.05.2003 & 72 & 70 & 0 & 0 & 0 & 0 & 0 &  0 \\
2452990.0--2453356.0 & 16.12.2003--16.12.2004 & 81 & 73 & 0 & 0 & 0 & 0 & 0 & 0 \\
\hline
\multicolumn{2}{r}{Total} & 381 & 361 & 179 & 248 & 92 & 88 & 38 & 128 \\
\hline
\end{tabular}
\end{table*}

   \begin{figure}[t]
   \centering
   \includegraphics[angle=0,width=8.7cm]{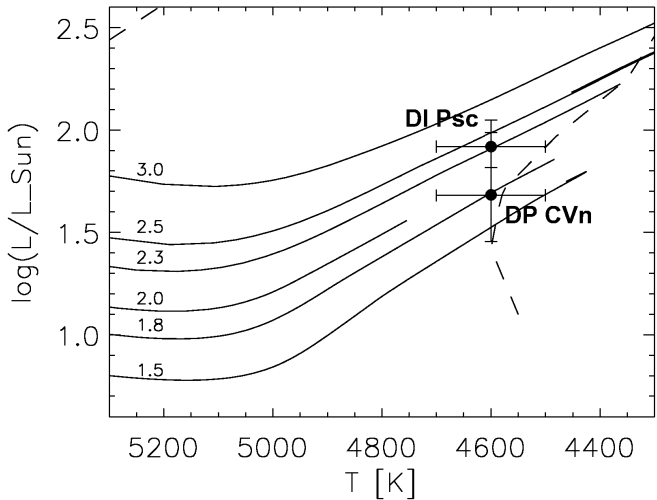}
   \caption{The observed positions of \dpcvn\ and \dipsc\ in the Hertzsprung-Russell diagram. The post-main-sequence tracks for 1.5, 1.8, 2.0, 2.3, 2.5 and
    3.0 solar masses are shown (c.f. Granzer et al. \cite{woozi1}). These tracks suggest masses of 1.8\,$M_{\odot}$ and 2.3\,$M_{\odot}$ for
    \dpcvn\ and \dipsc, respectively. Both stars appear on the RGB at the end of the first Li dredge-up phase, which is marked by the dashed lines
    taken from Charbonnel \& Balachandran (\cite{cha:bal}).}
   \label{F1}
   \end{figure}

\begin{figure}[tb]
\includegraphics[angle=-90,width=9cm]{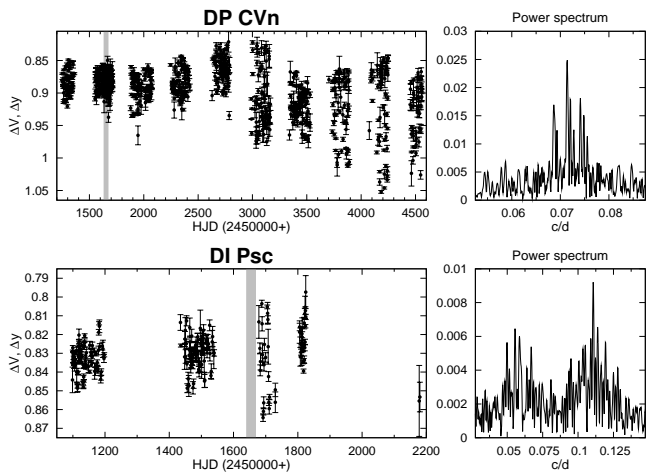}
\caption[ ]{$\Delta V$ and $\Delta y$ magnitudes and periodograms from the APT data of \dpcvn\ (top) and  \dipsc\ (bottom). The vertical grey stripes indicate
the time of our Doppler imaging data. The photometric periods adopted for this work are 14.010\,d (\dpcvn) and 18.066\,d (\dipsc, but see also Sect.~\ref{sect_rot}
for details). 
\label{F2}}
\end{figure}
\begin{figure}[tb]
\includegraphics[angle=0,width=9cm]{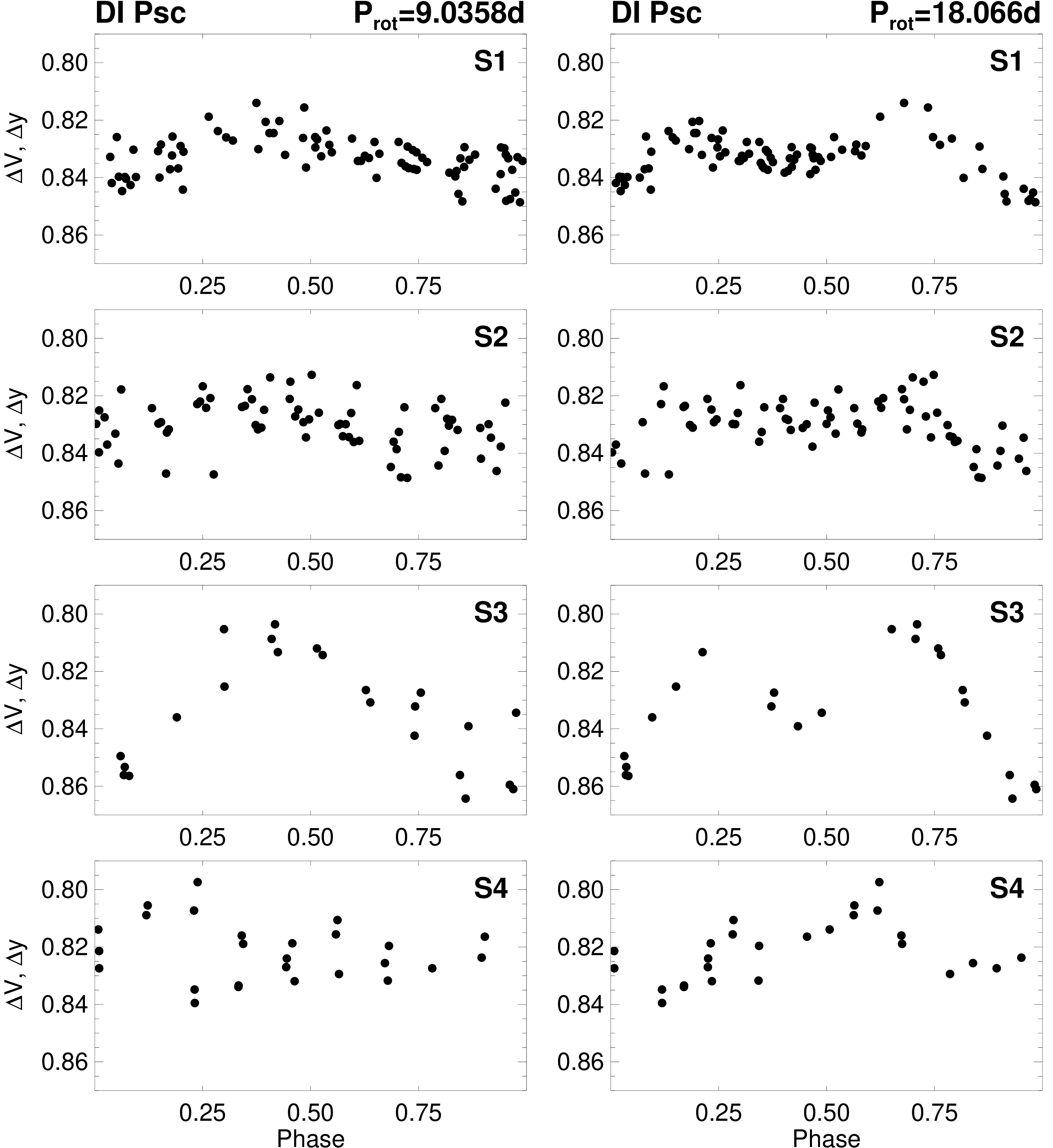}
\caption[ ]{Folded light curves of  \dipsc\  for four observing runs between HJD\,2451098.7 and HJD\,2451825.8 (S1-S4 from top to bottom panels, respectively),
assuming rotation period of either 9.0358\,d (left panels) or 18.066\,d (right panels), according to the power spectrum in Fig.~\ref{F2} bottom.
Note that using 18.066\,d yields less scattered light curves. For a quantitative comparison see Sect.~\ref{sect_rot}.
\label{F2.1}}
\end{figure}
\section{Stellar parameters}\label{sect_param}

\subsection{Absolute dimensions}

{\em Hipparcos} measured parallaxes for \dpcvn\ and \dipsc\ of 3.35$\pm$0.99~mas 
and 5.27$\pm$0.73~ mas, respectively (van Leeuwen~\cite{hip2}). This translates into distances of $299^{+125}_{-69}$ pc for \dpcvn\ and $190^{+30}_{-23}$ pc for
\dipsc. Table~\ref{T3} gives a summary of the astrophysical data of both systems.

The brightest observed $V$ and $I_{C}$ magnitudes for \dpcvn\  are 8$\fm$50 and 7$\fm$42 respectively, and for
\dipsc\ 7$\fm$10 and 5$\fm$95.  Due to the relatively large distances, interstellar reddening has to be taken into account
before using absolute magnitudes. The galactic coordinates of \dpcvn\  and \dipsc\ are $\ell$=128.61 $b$=+65.45 and
$\ell$=78.90 $b$=-47.96, respectively. We infer the reddening to these directions from the dust maps by Schlegel et~al. (\cite{schlegel})
and find $E(B-V)\approx$0\fm014 for \dpcvn\  and $E(B-V)\approx$0\fm073 for \dipsc. Using the transformations given by Schlegel et~al.\ this translates to
$A_{V}\approx 0\fm05$ and $A_{I_{C}}\approx 0\fm03$ for \dpcvn, and $A_{V}\approx 0\fm24$ and $A_{I_{C}}\approx0\fm14$
for \dipsc. Together with the {\em Hipparcos} distance this yields an absolute visual magnitude $M_{V}=+1\fm07_{-0.76}^{+0.57}$ and the
de-reddened $V-I_{C}$ color of 1$\fm$06 for \dpcvn, and $M_{V}=+0\fm47_{-0.32}^{+0.28}$ and $V-I_{C}$=1\fm05 for
\dipsc. It has to be noted that Schlegel et al.'s dust maps give the interstellar reddening throughout the entire Milky Way,
which means that the adopted reddening could be exaggerated in our case. Nevertheless, the target stars are located at such
distances and positions that they are likely above most of the dust, and the expected errors are well within the
errors caused by the uncertainties in other parameters.

The de-reddened $V-I_{C}$ color can be used to estimate the effective temperature of our stars, e.g., with the empirical formula
given by McWilliam (\cite{McW}) and the transformation of Bessell (\cite{bessell}). For both stars, we obtain $T_{\rm eff}$=4600$\pm$100\,K.
This is in agreement with an independent estimation from our $b-y$ data. The bluest observed $b-y$ colors are 0\fm70 and 0\fm71 for \dpcvn\ and \dipsc, respectively,
and, with the calibration of Clem et al. (\cite{clem}), suggest $T_{\rm eff}$=4550$\pm$100\,K for $\log g$=2.5 for both stars, in agreement
with the estimate from $V-I_{C}$ within its uncertainties. This gives a bolometric correction of --0.53 (Flower~\cite{flow})
and bolometric absolute brightness of $M_{\rm bol}=+0\fm54_{-0.76}^{+0.57}$ (\dpcvn) and $M_{\rm bol}=-0\fm06_{-0.32}^{+0.28}$ (\dipsc).
The logarithmic luminosity ($\log\frac{L}{L_{\odot}}$) of \dpcvn\ is $1.68_{-0.23}^{+0.30}$ and of \dipsc\ is $1.92_{-0.10}^{+0.13}$, based on an
absolute bolometric magnitude of the Sun of $M_{\rm bol,\odot}=4\fm 74$ (Cox \cite{cox}).

Morel et al. (\cite{morel}) determined temperatures from the excitation equilibrium of Fe\,{\sc i} lines of 13 active binaries and found only fair
agreement with the temperature from $B-V$ color, but even cooler by 200\,K on average from $V-R$ and $V-I$. Despite
that the temperatures from these three indices agree within their errors in case of our two active stars, we employed the line-ratio
calibration for giants by Strassmeier \& Schordan (\cite{str:sch}). This is normally not done because our Doppler-imaging stars have
rather strong rotational broadening that makes it virtually impossible to measure accurate line-ratios but may serve as an
independent cross check for the photometric temperatures. Altogether, eight line ratios between 6405\AA\ and
6460\AA\ were measured for both stars and, following the procedure suggested by Strassmeier \& Schordan
(\cite{str:sch}, their Fig.~18) for $V-I$ only, convert to average effective temperatures of 4530$\pm$75\,K and 4580$\pm$75\,K for
\dpcvn\ and \dipsc, respectively. Note that the \dpcvn\ line ratios appear 0-3\% \ larger than for \dipsc, which is reflected
in the different effective temperatures. However, the expected external uncertainties due to the large line broadening are at least
of the same order. We conclude that the line-ratio method at least agrees with the previous temperature determinations but fails to
improve the accuracy.

The positions of \dpcvn\ and \dipsc\ in the Hertzsprung-Russell (H-R) diagram are shown in Fig.~\ref{F1}. We have computed new post main sequence
evolutionary tracks with the Kippenhahn code with up-to-date input physics (as described in Granzer et al. \cite{woozi1}). No
overshooting was assumed though. The comparison with these tracks for solar metallicity suggests masses of approximately
$1.8_{-0.4}^{+0.7}$~$M_{\odot}$ (\dpcvn) and $2.3_{-0.3}^{+0.4}$~$M_{\odot}$ (\dipsc).
We placed our stars also on the colour-magnitude diagram constructed with the Padova stellar evolution code (Bertelli et al. \cite{padova})
(considering overshooting) which yielded slightly lower values but still within the given error boxes,
i.e., $1.7_{-0.3}^{+0.5}$~$M_{\odot}$ for \dpcvn\ and $2.1_{-0.2}^{+0.3}$~$M_{\odot}$ for \dipsc. 
This result places both stars quite high up on the red-giant branch (RGB) and very close to the end of the first dredge-up phase. \dpcvn\ has basically already
passed this phase. During the first dredge-up the Li abundance at the surface gets diluted as Li-free material from the stellar
interior is brought to the surface, i.e., both stars should have low Li abundance.

\begin{table}
\caption{Summary of astrophysical data. \label{T3}}
\begin{flushleft}
 \begin{tabular}{lll}
  \hline\noalign{\smallskip}
  Parameter                & \dpcvn                &  \dipsc \\
  \noalign{\smallskip}\hline\noalign{\smallskip}
  Distance (pc, Hipparcos)     & $299^{+125}_{-69}$ & $190^{+30}_{-23}$\\
  Spectral type            & K1III & K1III \\
  $M_{V}$              & +1\fm07$_{-0.76}^{+0.57}$ & +0\fm47$_{-0.32}^{+0.28}$\\
  Luminosity ($\log\frac{L}{L_{\odot}}$)         & $1.68_{-0.23}^{+0.30}$ & $1.92_{-0.10}^{+0.13}$ \\
  $\log g$ (adopted)       & 2.5 & 2.5 \\
  $V-I_{C}$ (dereddened)   & 1$\fm$06 & 1$\fm$05 \\
  $T_{\rm eff}$ (K)           & 4600$\pm$100    & 4600$\pm$100 \\
  $v\sin i$ (km\,s$^{-1}$)            & 39.0$\pm$1.0 & 42.0$\pm$1.0 \\
  Rotation period (d)   & 14.010$\pm$0.040 & 18.066$\pm$0.088 \\
  Inclination  ($\degr$)            & 35$\pm$10 & 50$\pm$10  \\
  Radius       ($R_{\odot}$)           & $18.7^{+7.3}_{-4.0}$  & $19.6^{+4.4}_{-2.7}$ \\
  Mass          ($M_{\odot}$)           & $1.8_{-0.4}^{+0.7}$  & $2.3_{-0.3}^{+0.4}$ \\
  LTE Li abundance (log)   & 2.28$\pm$0.18 &  2.20$\pm$0.15 \\
  \noalign{\smallskip}\hline
 \end{tabular}
\end{flushleft}
\end{table}

\subsection{Rotational properties}\label{sect_rot}

The rotation period of \dpcvn\ and \dipsc\ was determined from our photometric observations using the frequency analysis
program package {\sc MuFrAn} by Koll\'ath (\cite{mufran}), see Fig.~\ref{F2}.
For both of the stars $V$ and $y$ observations were used together, as this gave the largest amount of data points
for the analysis. The most significant period found in the \dpcvn\ observations was $14.010\pm 0.040$ days.
In the \dipsc\ data two similarly significant periods were found: 
$9.0358\pm0.060$\,d and $18.066\pm0.088$\,d.
Although, the shorter period is the more significant one, after a careful investigation the longer period was adopted.
The reasons for this are manifold: earlier period analysis (Strassmeier et al. \cite{kgs2}) indicates a period of 18.4\,d, two other period
analysis methods, the Lomb-method (Press et al. \cite{press92}) and the Three Stage Period Analysis (TSPA, Jetsu \& Pelt \cite{tspa}) also favour the
longer period. In addition, the individual light curves (S1-S4) in Fig.~\ref{F2.1} show less scatter when phasing them with the longer period.
For a quantitative test we apply spot model fits for the seasonal light curves ({\sc SpotModeL} by Rib\'arik et al. \cite{spot}) to compare
the sum of squares of residuals, assuming either $P_{\rm rot}$ of 9.0358\,d or 18.066\,d. Unambiguously, systematic
improvement of the fits is found when using the longer period, resulting in 53\% (S1), 30\% (S2), 89\% (S3) and 12\% (S4) less residuals.
Finally, we note that in a situation where there are two spot groups separated by approximately half a rotation, many period search methods
would prefer the period which is half of the real one.

Rotational phases were calculated using the following ephemerides:
\begin{equation}\label{eqdpcvn}
\rm{HJD}= 2,451,632.63084 + 14.010(40) \times E, \ \ \ (\dpcvn),
\end{equation}
\begin{equation}\label{eqdipsc}
\rm{HJD}= 2,451,639.00330 + 18.066(88) \times E, \ \ \ (\dipsc)
\end{equation}
where the epoch is the time of the first spectroscopic observation in the data set and the period is the photometric period. The errors
of the period determinations are estimated by increasing the residual scatter of the nonlinear least-squares solutions a certain degree,
which corresponds to 10\% of the accuracy of our photometric data (cf. Ol\'ah et al. \cite{oljust}).

Our Doppler imaging analysis (Sect.~\ref{sect_di}) helped us to determine the projected rotational velocities ($v\sin i$) and the
inclination angles $i$ from goodness of fit landscapes (see Fig.~\ref{F5}).
When we combine the resulting $v\sin i$ of 39.0$\pm$1.0 \kms\ for \dpcvn\ and 42.0$\pm$1.0 \kms\ for \dipsc\ with above rotational periods,
the minimum radii of our stars are
10.7$\pm$0.3\,$R_{\odot}$ and 15.0$\pm$0.4\,$R_{\odot}$, respectively. Doppler imaging constraints the inclination
to 35$\degr$$\pm$10$\degr$ for \dpcvn\ and 50$\degr$$\pm$10$\degr$ for \dipsc, giving absolute
radii of 18.7\,$R_{\odot}$ and 19.6\,$R_{\odot}$, respectively (see also Table~\ref{T3}). We note, that within the given errors, the radii and masses from
Table~\ref{T3} are in agreement with the adopted $\log g$ value.

\subsection{Lithium abundances}\label{sect_lithium}

\begin{figure}[tb]
\includegraphics[angle=0,width=9cm]{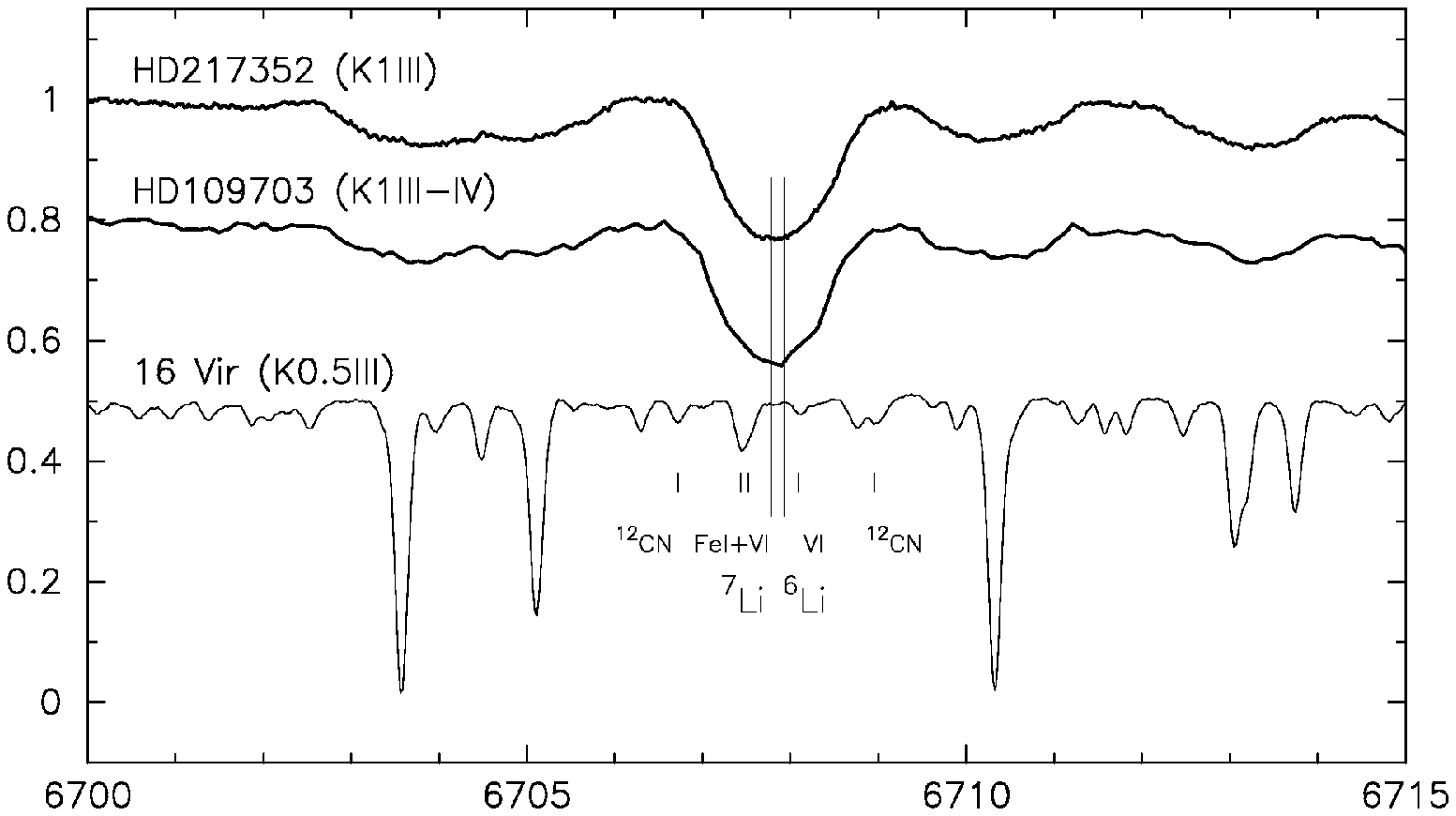}

\includegraphics[angle=0,width=4.45cm]{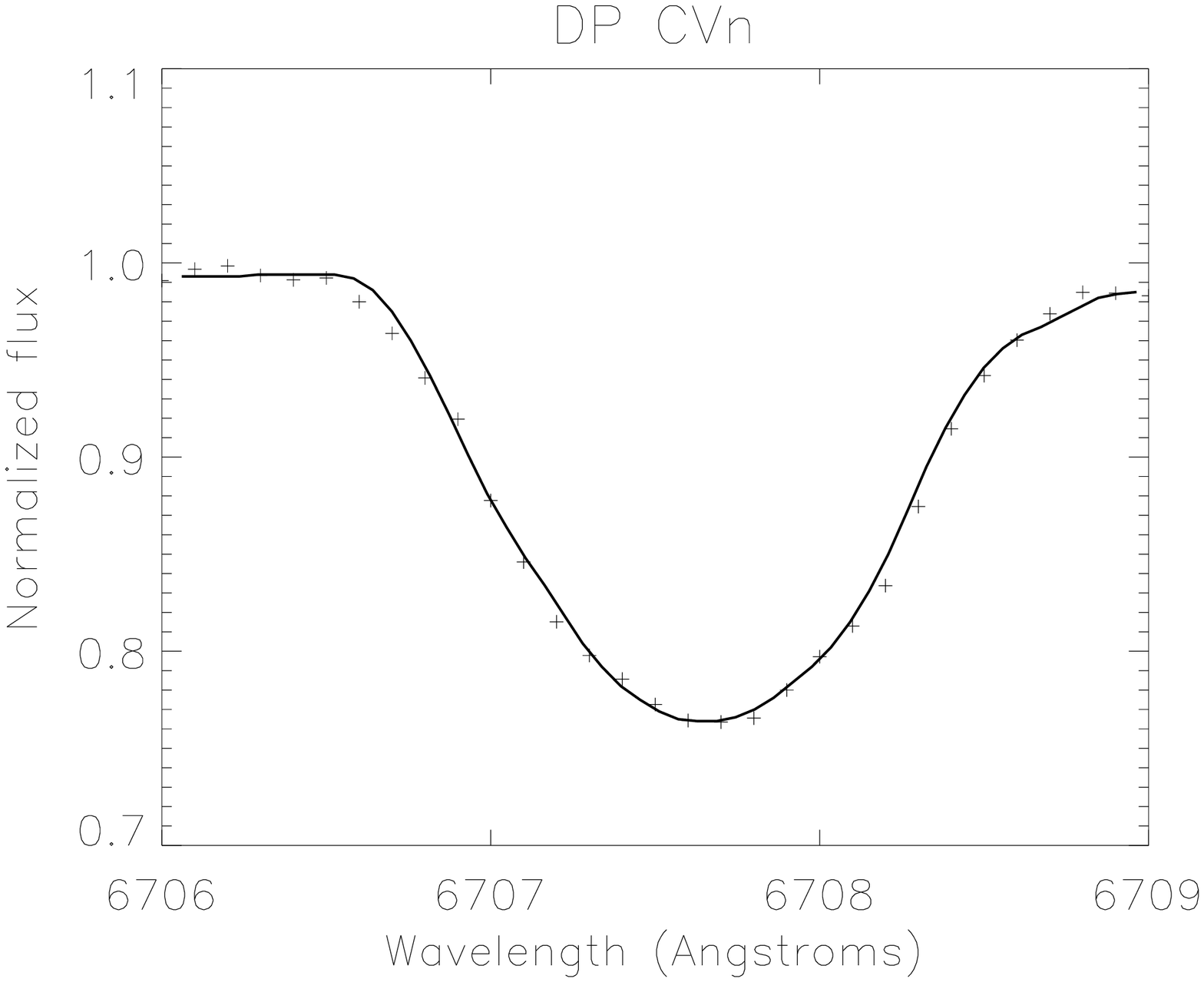}\vspace{0.1cm}\includegraphics[angle=0,width=4.45cm]{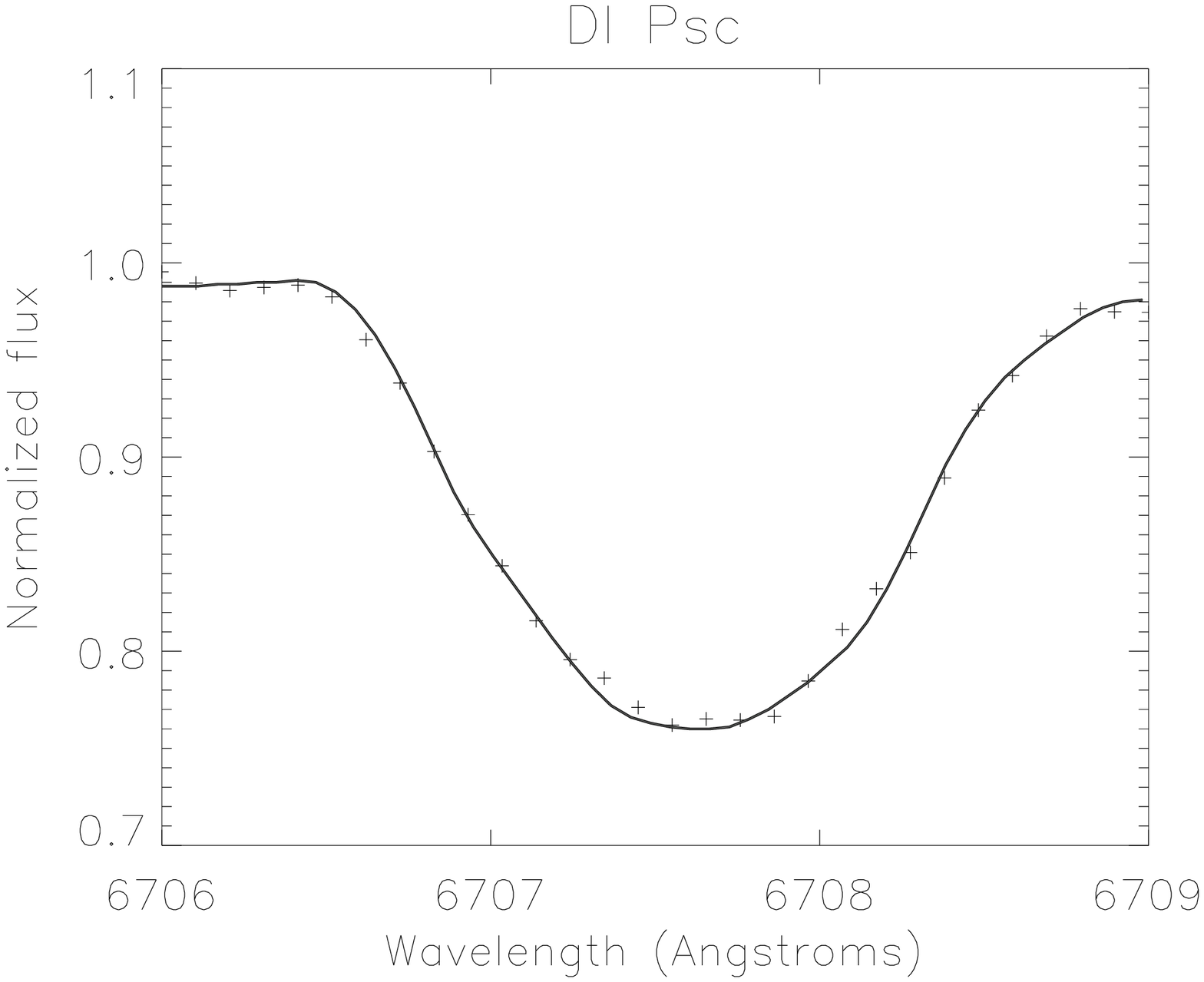}
\caption[ ]{{\em Top panel:} Li\,{\sc i} 6708\,\AA\ spectra of \dipsc\ (top) and
\dpcvn\ (shifted by --0.2). Also shown is a spectrum of the inactive K0.5III giant 16\,Vir (shifted to bottom) for comparison purpose.
The measured equivalent width of the lithium line of \dipsc\ and \dpcvn\ is 328$\pm$5~m\AA \ and 316$\pm$7\,m\AA , respectively.
{\em Bottom panels:} Li\,{\sc i} 6708\,\AA\ line fitted by synthetic spectra for \dpcvn\ (left) and \dipsc\  (right), resulting in logarithmic abundances of $\log n$(Li)=2.28, and of 2.20, respectively. \label{F3}}
\end{figure}

Top panel of Fig.~\ref{F3} shows a high-resolution spectrum of the exceptionally strong Li\,{\sc i} 6708\AA\ line for both stars in comparison
with the inactive K0.5III star 16\,Vir (otherwise an IAU velocity standard; see, e.g., Strassmeier et al. \cite{cmcam},
Weber \& Strassmeier \cite{web:str}, Strassmeier et al. \cite{straetal11}, etc.).
The measured equivalent widths are 316$\pm$7\,m\AA \ (\dpcvn) and 328$\pm$5\,m\AA\ (\dipsc).
The corresponding equivalent width ($EW$)
of the sum of the lines that would
appear as blends within the Li line width, as measured in 16\,Vir, is 39.8$\pm$3\,m\AA. The uncertainty of just 5-7\,m\AA\ is estimated
from different types of fits to the line profile, e.g. with a
Gaussian profile of variable width where only the profile points of the red profile wing are used for the least-squares fit, from a normal
Gaussian fit or from a simple profile-area integration. The
final Li $EW$s for the two target stars are obtained by subtracting the 16\,Vir contribution and are 277\,m\AA\ for \dpcvn\ and 288\,m\AA\ for \dipsc.

Next we fit synthetic (LTE) spectra to the Li line profiles. Synthetic spectra were calculated and matched to the observed spectra by adjusting
the Li abundance to fit the 6707.8\,\AA\
feature. We used the VALD database (Piskunov et al. \cite{pisk3}, Kupka et al. \cite{kupka}) and Kurucz's CD-ROM No. 23 for the
numerous atomic and molecular lines. Its $\log gf$ values were slightly tuned
using a spectrum of $\beta$\,Gem, while we adopted (and fixed) the values from Reddy et al. (\cite{red:lam1}) with the component structure
of the Li resonance doublet from Hobbs et al. (\cite{hobbs}).
We assume that Li is purely $^7$Li.

Bottom panels of Fig.~\ref{F3} show comparisons of the observed and synthetic spectra for both stars.
Li abundances derived from our LTE analysis are equal to 2.28 for \dpcvn\ and 2.20 for \dipsc. 
Small perturbations of the observed profiles are due to spots on the stellar surfaces which may influence the abundance value. Note that 
100\,K uncertainty in $T_{\rm eff}$ yields $\approx$0.2\,dex.  

Tests with changes of the continuum level of 2\% resulted in abundance changes of 0.10--0.12\,dex. In our case the continuum level
of the observed spectra was obtained iteratively from a comparison with calculated spectra including  rotationally broadening.  Notice a 4\% \
depression of the continuum in the bottom panels of Fig.~\ref{F3} due to the rotational convolution with
several absorption blends.
We did not carry out a full abundance analysis and thus assumed solar abundances for all other elements except Li and C,N,O.
As our target stars likely belong to the ``bump'' clan (Charbonnel \& Balachandran \cite{cha:bal}) corresponding values of C,N,O abundances
were used in our calculations.

Non-LTE corrections interpolated from the results by Carlsson et al. (\cite{car:rut}) were found to be small ($<$0.03\,dex) and
comparable to its uncertainties and methods of interpolation. Therefore, we refrain from giving non-LTE abundances. However, we
note that the non-LTE curves of growth of Pavlenko \& Magazz\'u (\cite{pav:mag}) for a 4600\,K/$\log g$=2.5 model convert the
$EW$s measured into a Li abundance of 2.37$\pm$0.05 (\dpcvn) and 2.40$\pm$0.05 (\dipsc), both on the $\log n$(H)=12.00 scale,
i.e., larger by 0.1-0.2\,dex than our LTE
values.

The position in the H-R diagram in Fig.~\ref{F1} and the large Li abundance places both stars at the end of the first dredge up
according to the models of Charbonnel \& Balachandran
(\cite{cha:bal}). Thus, they are truly Li-rich stars. They even clump together with stars having freshly synthesized Li on their surface,
such as the super-meteoritic Li stars HD\,9746 and
HD\,233517 (Fekel \& Balachandran \cite{fek:bal}).

\begin{figure}[t]
\begin{center}
\includegraphics[angle=0,width=4.5cm]{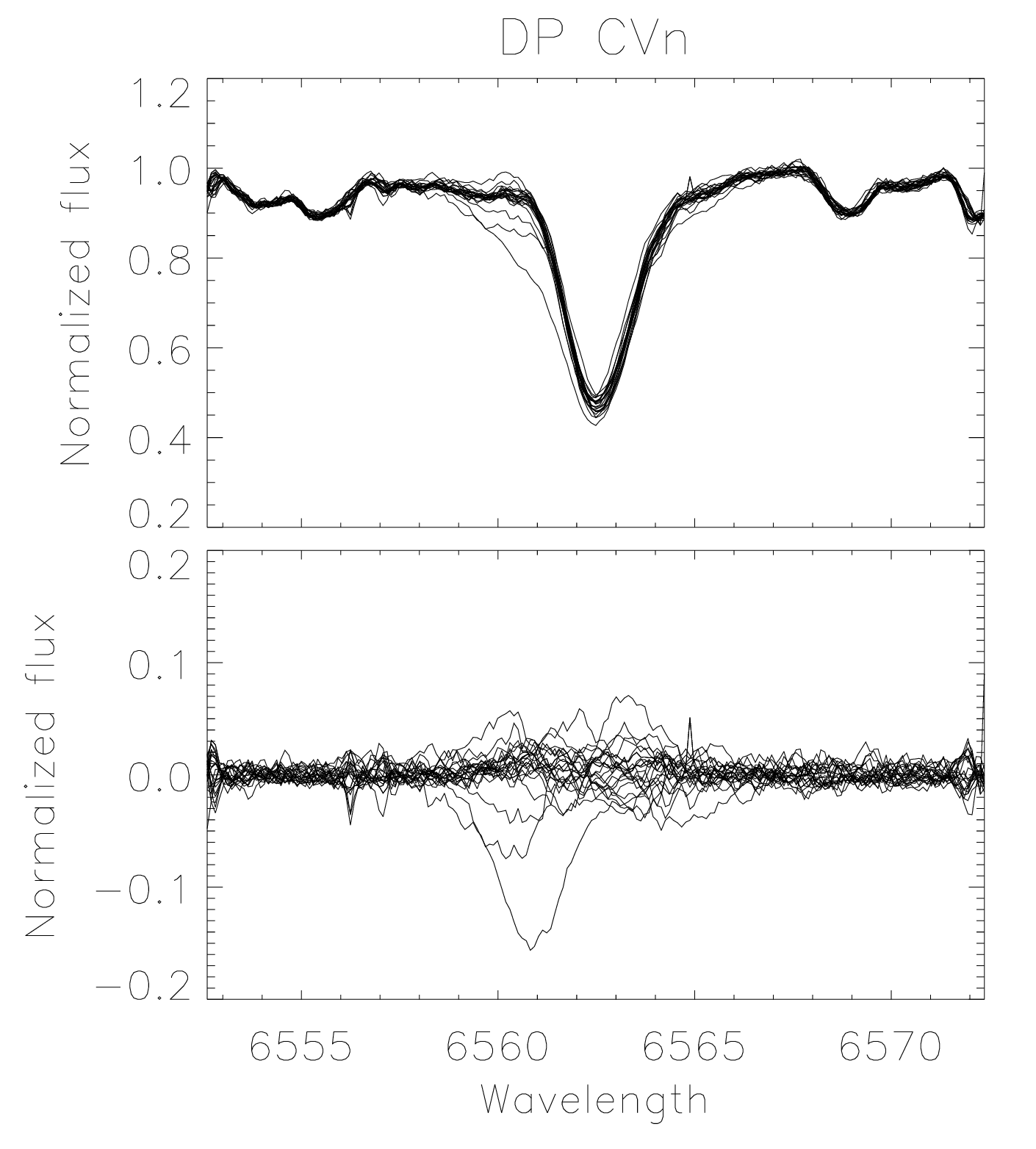}\includegraphics[angle=0,width=4.5cm]{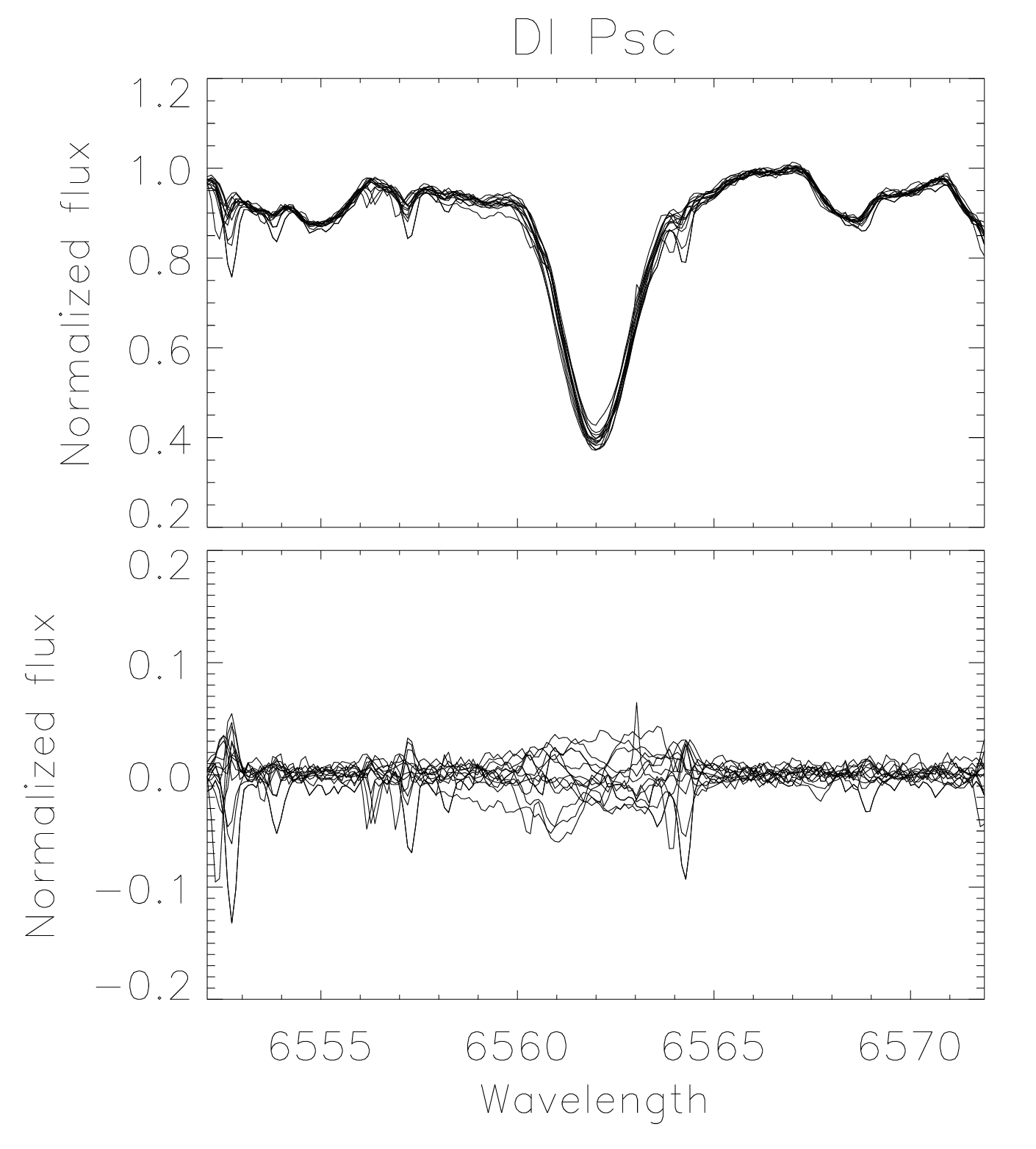}
\end{center}
\caption[ ]{\Halpha line profiles of \dpcvn\ (left) and of \dipsc\ (right). In the top panels the observations are plotted together,
while in the bottom panels the corresponding averages are subtracted to emphasize the nightly variations.  Note the telluric water lines, especially
at around 6552.5\,\AA,  6557\,\AA, and 6564\,\AA,
as well as the sporadic cosmic peaks.
\label{F4}}
\end{figure}

\subsection{H$\alpha$ line profiles}\label{sect_halpha}

In both stars \Halpha appears as a relatively stable absorption line without periodic variations from one spectrum to the next (see Fig.~\ref{F4}),
suggesting a rather homogeneous
chromosphere made up of symmetrically distributed emission regions, comparable to a more active version of the solar chromospheric
network. The averaged $EWs$ are 1.183\,\AA\ for \dpcvn\ and 1.369\,\AA\ for \dipsc, virtually without any significant phase dependency.
The only markable phase independent change is seen in three spectra of \dpcvn\ taken in three subsequent nights
from HJD 2,451,636.657 ($\phi$=0.288) through HJD 2,451,637.745 ($\phi$=0.365, with $EW$ of 1.526 at maximum) to HJD 2,451,638.698 ($\phi$=0.433).
The line is broadened asymmetrically towards the blue wing, however, the feature did not return during the next rotation cycle. Such a non-recurring
phenomenon can be caused by an erupting coronal prominence. A similar change in the \Halpha line was reported e.g., for the effectively single
G8-giant CM\,Cam by Strassmeier et al. (\cite{cmcam}).

\section{Doppler imaging}\label{sect_di}

\subsection{The {\sc TempMap} inversion code}

A full description of the {\sc TempMap} code can be found in Rice et al. (\cite{rice1}), Rice \& Strassmeier (\cite{tests}) and
in Rice (\cite{rice2}). The program performs a full LTE spectrum synthesis by solving the equation of transfer through a set
of ATLAS-9 (Kurucz \cite{kur}) model atmospheres at all aspect angles and for a given set of chemical abundances. Atomic line
parameters are obtained from VALD (Piskunov et al. \cite{pisk3}, Kupka et al. \cite{kupka}). Spectra were calculated for temperatures
ranging from 3500\,K to 5500\,K in steps of 250\,K and with solar abundances. Simultaneous inversions of the usual mapping lines
as well as of the photometric light variations in two photometric bandpasses are then carried out using maximum-entropy regularization.

\begin{figure}[b!!]
\includegraphics[angle=0,width=4.5cm]{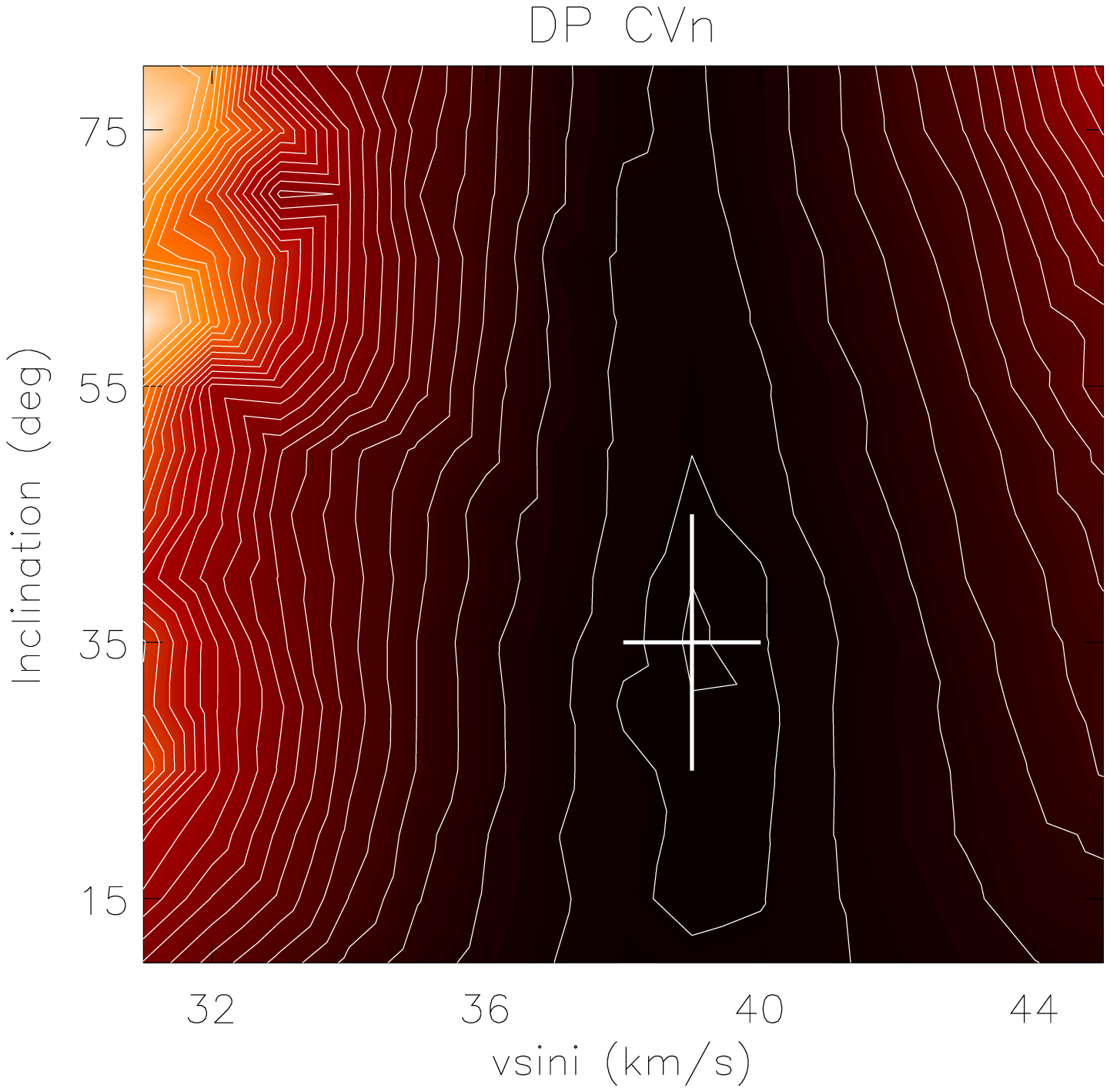}\includegraphics[angle=0,width=4.5cm]{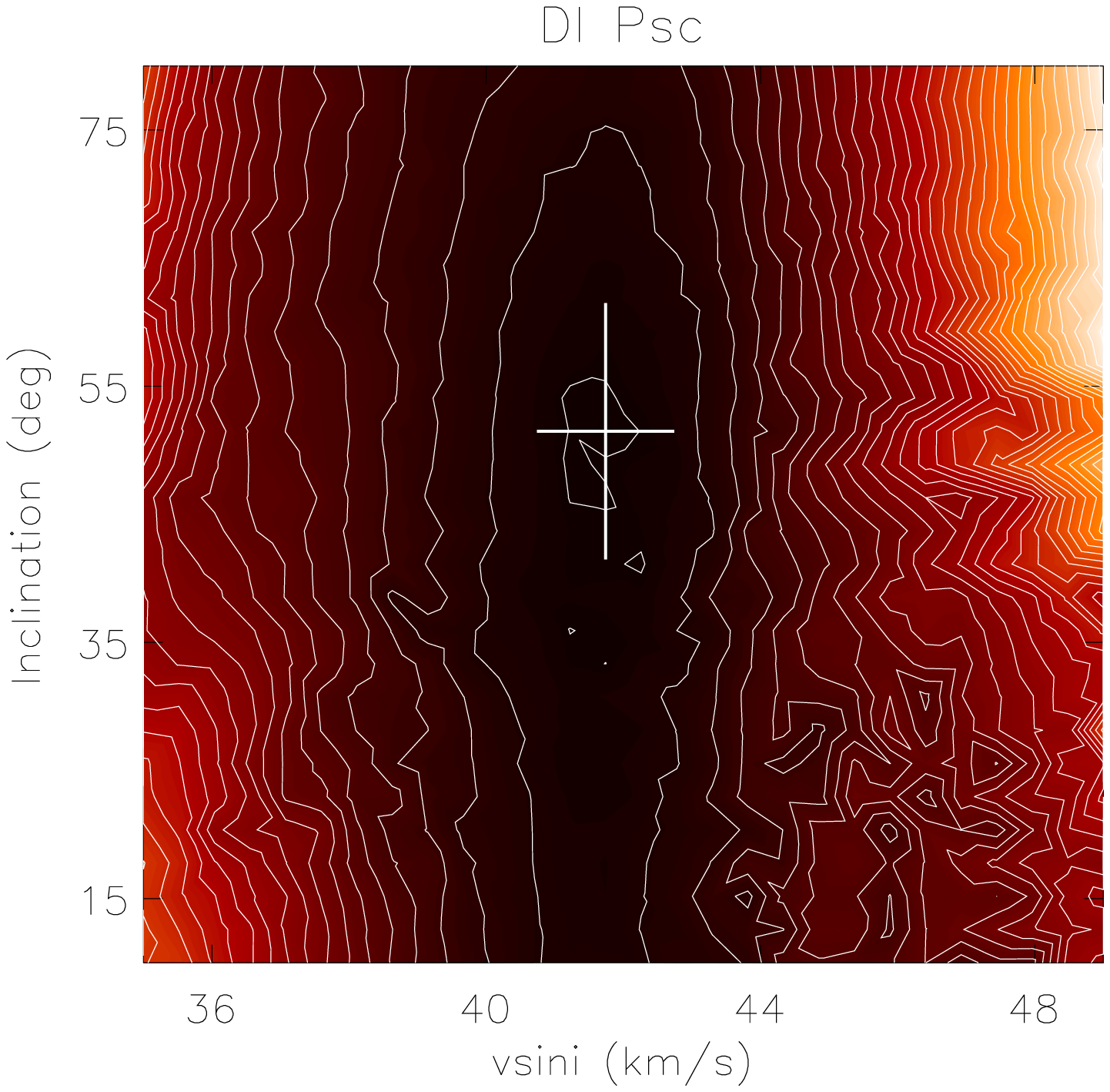}
\caption[ ]{The dependency of the normalized goodness of fit ($\chi^2$) of our Doppler maps over the $v\sin i - i$ parameter plane.
Best values for \dpcvn\ (left) are $v\sin i$ of 39.0$\pm$1.0\,km\,s$^{-1}$ with $i\approx35$$\degr$$\pm$10$\degr$
and for \dipsc\ (right) are 42.0$\pm$1.0\,km\,s$^{-1}$ with 50$\degr$$\pm$10$\degr$, respectively.
\label{F5}}
\end{figure}

A large number of trial inversions helped us to fine tune some input parameters, and led us to the final values
for the projected rotational velocity $v_{\rm eq}\sin i$ and the inclination $i$, which were determined with test runs by minimizing the $\chi^2$, i.e., the goodness-of-fit
of the Doppler reconstructions (for the $\chi^2$ definition see Rice \& Strassmeier \cite{tests}). 
The goodness-of-fit landscape in Fig.~\ref{F5} demonstrates the change of $\chi^2$ over a meaningful domain of the $v_{\rm eq}\sin i - i$ parameter plane for both stars.
Best values for $v_{\rm eq}\sin i$ were 39.0$\pm$1.0\,km\,s$^{-1}$ (\dpcvn)
and 42.0$\pm$1.0\,km\,s$^{-1}$ (\dipsc), and for the inclination were  35$\degr\pm$10$\degr$ (\dpcvn) and 40$\degr\pm$10$\degr$ (\dipsc).
The best combinations are listed also in Table~\ref{T3}.

\begin{figure*}[t!!!!!!!!]
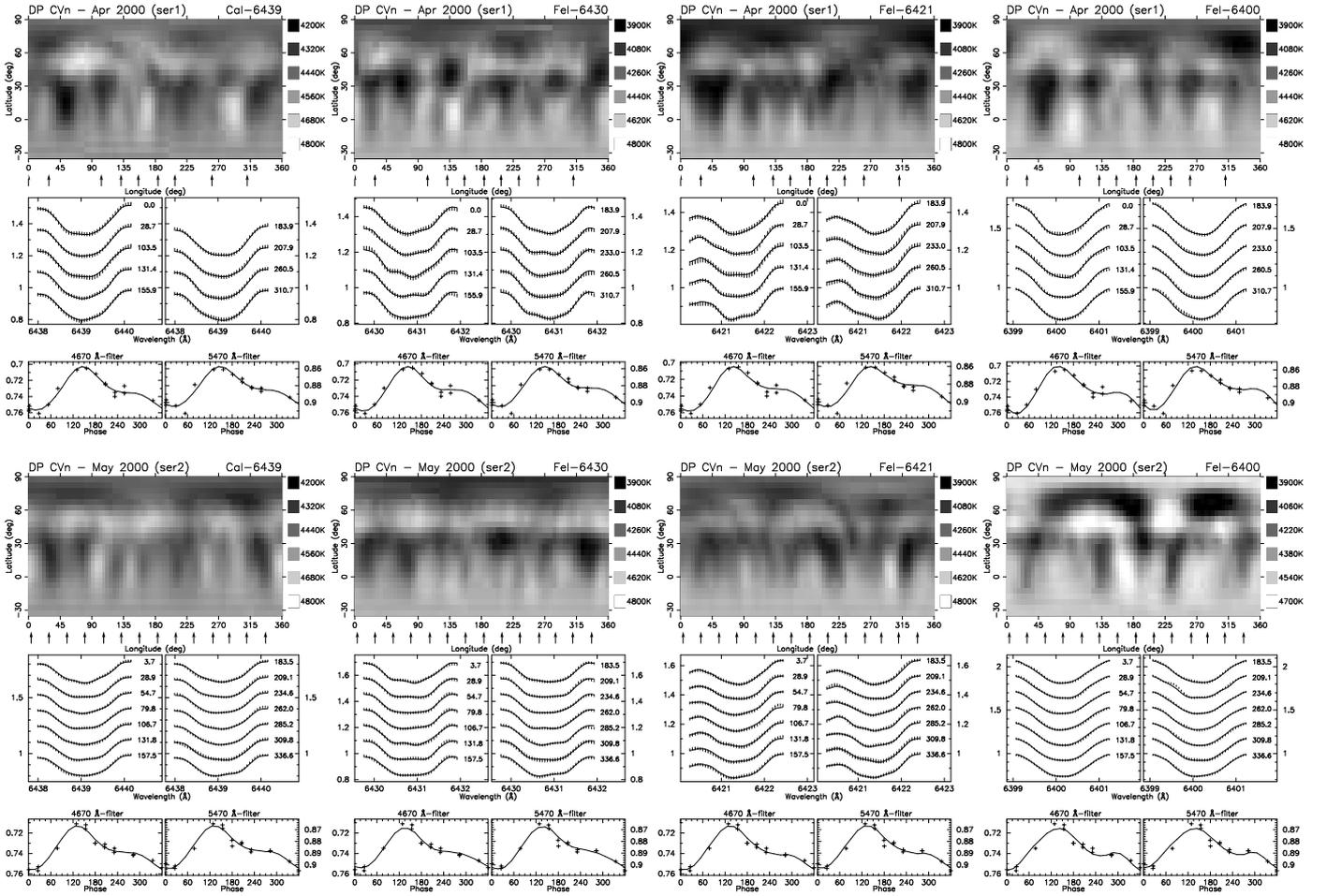

\includegraphics[angle=0,width=4.5cm]{F6_dpcvn_ca6439_s1.eps} \includegraphics[angle=0,width=4.5cm]{F6_dpcvn_fe6430_s1.eps} \includegraphics[angle=0,width=4.5cm]{F6_dpcvn_fe6421_s1.eps} \includegraphics[angle=0,width=4.5cm]{F6_dpcvn_fe6400_s1.eps}
\\
\\
\includegraphics[angle=0,width=4.5cm]{F6_dpcvn_ca6439_s2.eps} \includegraphics[angle=0,width=4.5cm]{F6_dpcvn_fe6430_s2.eps} \includegraphics[angle=0,width=4.5cm]{F6_dpcvn_fe6421_s2.eps} \includegraphics[angle=0,width=4.5cm]{F6_dpcvn_fe6400_s2.eps}
\caption[ ]{Surface temperature distribution maps of \dpcvn\  for the two available data subsets taken in April 2000 (ser1, top row) and in May 2000 (ser2, bottom row),
using Ca\,{\sc i} 6439\,\AA, Fe\,{\sc i} 6430\,\AA, Fe\,{\sc i} 6421\,\AA, and Fe\,{\sc i} 6400\,\AA\ mapping lines. The temperature maps are presented in a
pseudo-Mercator projection from latitude $-$35\degr\ to +90\degr. The phases of the observations are marked by arrows underneath.
Below surface temperature reconstructions, in the middle panels the fitted line profiles are shown.
Small dashes represent the data points and measure the $\pm$1-$\sigma$ errors.
Simultaneous light curves in Str\"omgren $b$ and $y$ are plotted in the bottom panels with their respective fits.
\label{F6}}
\end{figure*}

\subsection{Doppler reconstructions for \dpcvn}\label{sect_DI_dpcvn}

Data for this star were taken in the modus of 12 nights on, 16 nights off, 13 nights on, i.e., approximately two rotation cycles were covered by the observations,
with one rotation between them. This enabled two independent data sets with one spectrum per night and a total of 10 spectra for the
first rotation (ser1, note that one Ca\,{\sc i} 6439\,\AA\ spectrum was omitted due to probable cosmic noise) and 14 for the second rotation (ser2).
Four nights were lost during the first rotation due to bad weather, the second rotation was covered completely.

The sampling of our spectra allowed $\approx$23 pixels across the full width of a given spectral line and the S/N per pixel is between 100:1
to 150:1 as measured with IRAF. Additionally, there are 17 simultaneous two-colour photometric data points in
Str\"omgren $b$ and $y$.

\begin{figure}[th!!!!]
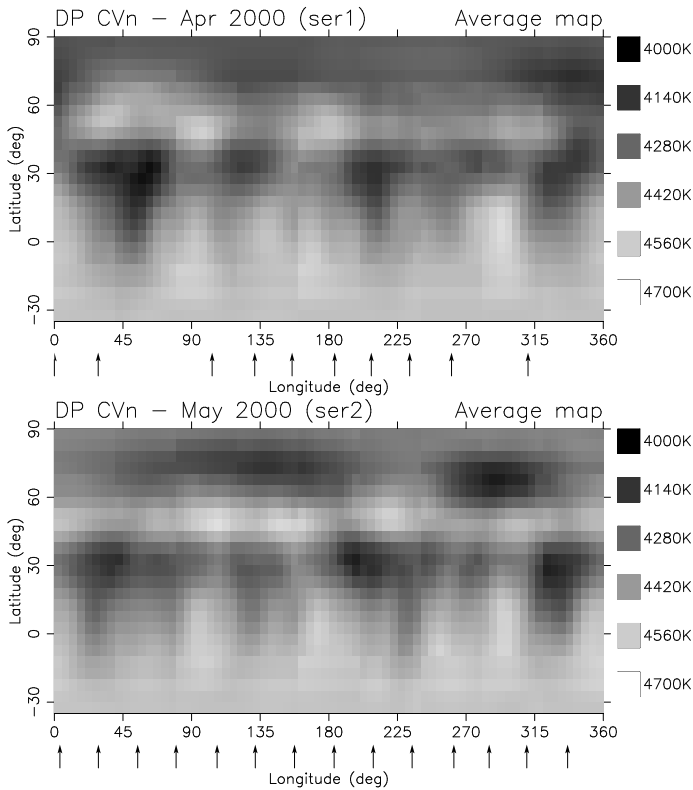

\includegraphics[angle=0,width=9.0cm]{F7_dpcvn_avgmap_s1.eps}
\\
\includegraphics[angle=0,width=9.0cm]{F7_dpcvn_avgmap_s2.eps} 
\caption[ ]{Combined images of \dpcvn\ for the two data subsets ser1 (top) and ser2 (bottom). The
maps are presented in a pseudo-Mercator projection from latitude $-$35\degr\ to +90\degr. The phases of the observations are marked by arrows below the maps.
Note, that even if the individual reconstructions show differences (cf. Fig.~\ref{F6}), the average maps still retain most of the traits. 
\label{F7}}
\end{figure}

Maps for the two data subsets taken in Apr 2000 (ser1) and in May 2000 (ser2) are
shown in Fig.~\ref{F6} for Ca\,{\sc i} 6439\,\AA, Fe\,{\sc i} 6430\,\AA,  Fe\,{\sc i} 6421\,\AA,
and Fe\,{\sc i} 6400\,\AA, our most commonly used absorption lines with well-known formation physics and mostly free of telluric blends.
(Surface reconstruction from Li can not be done because we have only one high-resolution Li spectrum.)
Also shown are the simultaneously fitted light curves {\bf in $b$ and $y$.}
The individual reconstructions agree quite well, revealing dominant spots at lower latitudes with  $\approx$700\,K below $T_{\rm eff}$, as well as some
high-latitude features 
but mostly with lesser contrast.
Note also, that the minimum temperature value of the Ca-maps in Fig.~\ref{F6} is set to 4200\,K, i.e., 300\,K hotter than that of the Fe-maps.
The less contrast of the Ca-line reconstruction
is probably due to the fact that the Ca line has broader equivalent width and also is more sensitive to temperature changes than the iron lines.
On the other hand, smaller differences
could occur through the relatively low S/N of the spectra. The agreement of the individual reconstructions can be demonstrated
by comparing them with their averages (see Fig.~\ref{F7}). Markedly, most of the noticeable features in the contemporaneous individual maps in Fig.~\ref{F6}
can be identified in the corresponding combined map.

\subsubsection{Searching for surface differential rotation}\label{sect_ccf}

Latitude dependency of the spot displacements from the first subset (ser1) in Apr 2000 to the second one (ser2) in May 2000 can be assigned 
as a trace of surface differential rotation.
Such a dependency is suspected when looking at Fig.~\ref{F6}, e.g., where the spots near the equator
seem to be shifted mostly backwards in longitude, unlike spots at higher ($>$45$\degr$) latitudes.
Differential rotation can be measured by cross-correlating subsequent Doppler images (Donati \& Collier Cameron \cite{doncol97}). 
We apply the cross-correlation method called ACCORD,
as described recently in K\H{o}v\'ari et al. (\cite{zetandr}). We cross-correlate the corresponding longitude strips of the subsequent Doppler maps for each latitude value
(in practice in bins of 5$\degr$ width) for the four available individual reconstructions. This way, we get four cross-correlation function (CCF) maps.
Then, the CCF maps are combined to achieve the average correlation pattern, wherein the sign of the differential rotation is expected
to be intensified as a common feature of the
individual CCF maps. Finally, the best correlated pattern is fitted by a solar-like differential rotation law, assuming
the usual quadratic form of $\Omega(\beta)=\Omega_{\rm eq}-\Delta\Omega\sin^2\beta$,
where $\Omega_{\rm eq}$ is the angular velocity at the equator and $\Delta\Omega=\Omega_{\rm eq}-\Omega_{\rm pole}$ is the
difference between the equatorial and the polar angular velocity.
The dimensionless surface shear parameter is expressed as $\alpha=\Delta\Omega/\Omega_{\rm eq}$.
\begin{figure}[th!!!!!!!!!!!!]
\includegraphics[angle=0,width=9.0cm]{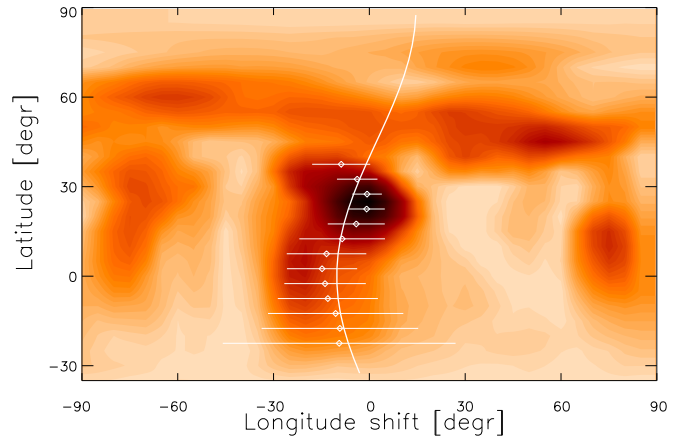}
\caption[ ]{Cross-correlation function (CCF) map for \dpcvn. The image is the average of four CCF maps obtained from cross-correlating the individual
Doppler reconstructions (i.e., Ca\,{\sc i} 6439\,\AA, Fe\,{\sc i} 6430\,\AA,  Fe\,{\sc i} 6421\,\AA, and Fe\,{\sc i} 6400\,\AA\ maps)
taken in Apr (ser1) and May (ser2). Dots with error bars are
the Gaussian-fitted correlation peaks per 5$\degr$ latitude bin.
The best fit suggests an anti-solar differential rotation function (solid line) and  corresponds to a surface shear $\alpha$ of $-$0.035.
\label{F8}}
\end{figure}

In Fig.~\ref{F8} the best-fit rotation law to the correlation peaks per latitude bin (represented by fitted Gaussian maxima and their errorbars) is plotted.
The relevant part of the correlation pattern extends
from $-$25$\degr$ to $+$40$\degr$, and is strongest at around 30$\degr$, i.e., the latitude belt that rotates with the disk-integrated photometric period.
This is not surprising because below
the equator the inclination angle limits the visibility while at higher latitudes the lack of strong surface features that
are jointly present on the corresponding maps yields weak
and slurred correlation. Note that at even higher latitudes, close to the visible pole, the geometric singularity makes ``correlation'' comparably meaningless.
The resulting fit indicates antisolar-type differential rotation, i.e. equatorial deceleration, with $\Omega_{\rm eq}=25.336\pm 0.044\degr$/day
and $\Delta\Omega=0.875\pm0.411\degr$/day or $\alpha=-0.035\pm0.016$ (errors are estimated from the FWHMs and amplitudes of the Gaussian profiles).
We note, that considering the moderate significance, the surface shear parameter needs further validation from better quality data.

\subsection{Doppler reconstructions for \dipsc}

\begin{figure*}[t!!!!]
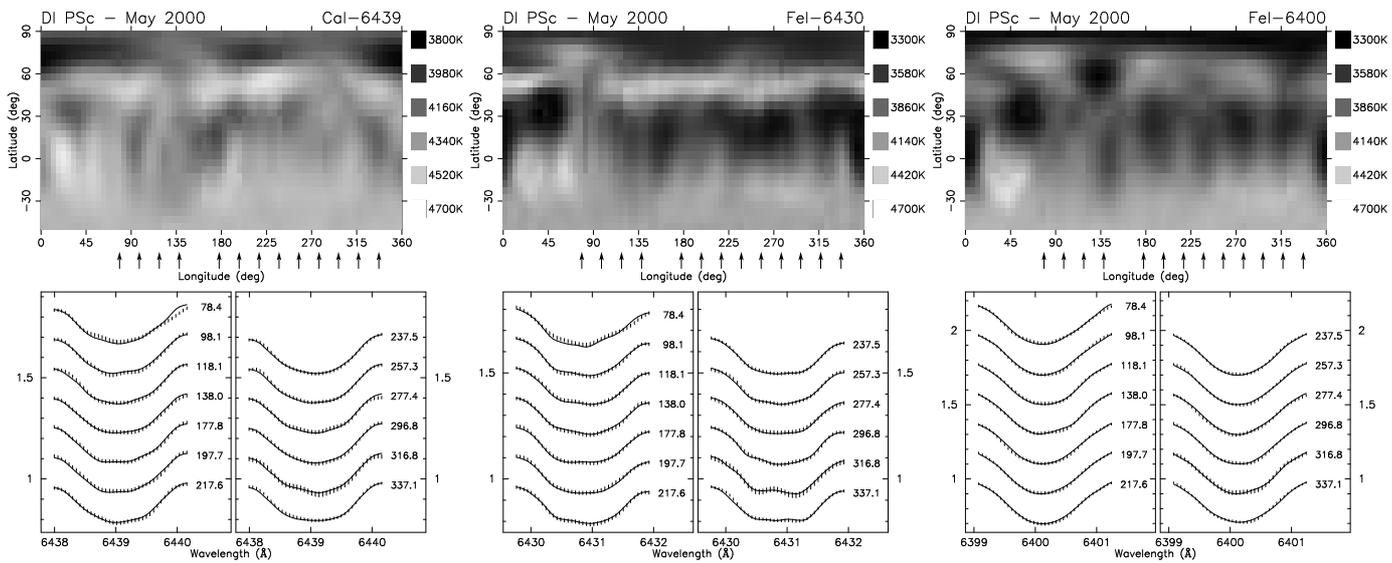

\includegraphics[angle=0,width=6cm]{F9_DIPSC_ca1_6439.eps} \includegraphics[angle=0,width=6cm]{F9_DIPSC_fe1_6430.eps} \includegraphics[angle=0,width=6cm]{F9_DIPSC_fe1_6400.eps}
\caption[ ]{Reconstructed Doppler maps of \dipsc\ for Ca\,{\sc i} 6439\,\AA, Fe\,{\sc i} 6430\,\AA, and Fe\,{\sc i} 6400\,\AA\ mapping lines. The
maps are presented in a pseudo-Mercator projection from latitude $-$50\degr\ to +90\degr. The phases of the observations are marked by arrows underneath.
Below surface temperature reconstructions the fitted line profiles are shown, where small dashes represent the data points and measure the $\pm$1-$\sigma$ errors.
\label{F9}}
\end{figure*}

Our observations allow for one Doppler reconstruction from data taken between HJD 2,451,661.00 and HJD 2,451,673.99, i.e. covering
13 nights, or 72\% of one rotation cycle.
Unfortunately, no simultaneous photometric data were available (but see Strassmeier et al.~\cite{LQHya1} on the role of photometry in {\sc TempMap}).
Fig.~\ref{F9} shows the Ca\,{\sc i} 6439\,\AA, Fe\,{\sc i} 6430\,\AA, and Fe\,{\sc i} 6400\,\AA\ images along with the corresponding
spectroscopic data and their respective fits.

Attempts to use other mapping lines from the observed spectral region, i.e. Fe\,{\sc i} 6421\,\AA, Fe\,{\sc i} 6411\,\AA\ or Fe\,{\sc i} 6393\,\AA\ led to
unacceptable results probably due to unknown blends (atomic and molecular), to uncertain turbulence parameters, etc.
Note that the available Li spectra of \dipsc\ are insufficient for Doppler reconstruction.

\begin{figure}[b!!!!]
\includegraphics[angle=0,width=9cm]{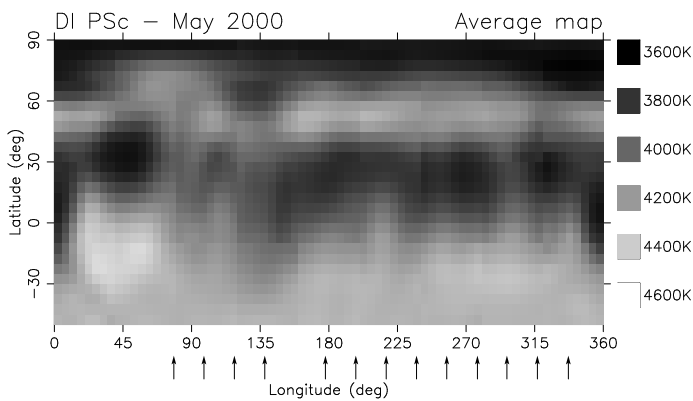}
\caption[ ]{Combined image of \dipsc\ from the available three individual Doppler reconstructions. The
map is presented in a pseudo-Mercator projection from latitude $-$50\degr\ to +90\degr. The phases of the observations are marked by arrows below the map.
\label{F10}}
\end{figure}

The Ca\,{\sc i} 6439\,\AA\ map reveals cool spots at high latitudes as well as spots at lower latitudes, but with lesser contrast.
Similarly, the Fe\,{\sc i} 6430\,\AA, and Fe\,{\sc i} 6400\,\AA\ maps show cool polar features, but dominant spots appear also at latitudes
below $\approx$40$\degr$, forming a belt-like structure. The temperature scale is different, typically ranging
from $\Delta T\approx 800$\,K in the Ca map up to 1300\,K in the iron maps. Aside from that and from minor
dissimilarities (most likely due to the moderate quality of the data and the uncovered phase gap from $\approx$340\degr\ to $\approx$80\degr),
the agreement is still acceptable, as shown
in Fig.~\ref{F10} where an average map is combined from the individual reconstructions. It is clearly worth to
revisit this star with instruments with larger wavelength coverage and further explore its surface inhomogeneity. 

\section{Long-term photometry of \dpcvn}\label{sect_longphot}

\subsection{Spot evolution}\label{subsect_spotevo}

The spot evolution was investigated first by plotting the maximum, minimum and mean magnitudes in $V$ and $y$ at different observing periods against
the amplitude. No systematic trend in the mean magnitude with amplitude is seen, while the maximum and minimum magnitudes are
systematically higher and lower, respectively, when the amplitude increases. We conclude that on \dpcvn\ the light curve variations are
caused by a rather symmetrical rearrangement of a more or less constant amount of dark spots.
This is consonant with the results from Doppler imaging (Sect.~\ref{sect_DI_dpcvn}), where
the most dominant spots are located within a belt below $\approx$40$\degr$ and distributed more or less evenly (see Fig.~\ref{F7}).

\subsection{Active longitudes}

The photometric dataset of \dpcvn\ is suitable for carrying out a more detailed analysis of the spot evolution through the 6 year-long observing range.
For this purpose the $V$ and $y$ observations were divided into 16 subsets. These individual light-curves were then
used for obtaining spot filling factor maps of the stellar surface. For the inversions we use the Occamian approach 
method (Berdyugina \cite{ber1}, but see also Lanza et al. \cite{lanza}). The stellar surface was divided into areas of $10\degr\times
10\degr$. The temperatures for the unspotted surface and the spots were fixed to 4600\,K and 4000\,K, respectively, based on the
Doppler maps (see Sect.~\ref{sect_DI_dpcvn}). The limb-darkening coefficient for the $V$ band for the unspotted surface was
estimated to be 0.802 from the results by Al-Naimiy (\cite{aln}) and the same value was also adopted for the spotted areas. As unspotted magnitude
we used $8\fm45$ as an estimation from the brightest observed magnitude of $8\fm50$.
\begin{figure*}[t!!!!!]
\includegraphics[width=4.5cm]{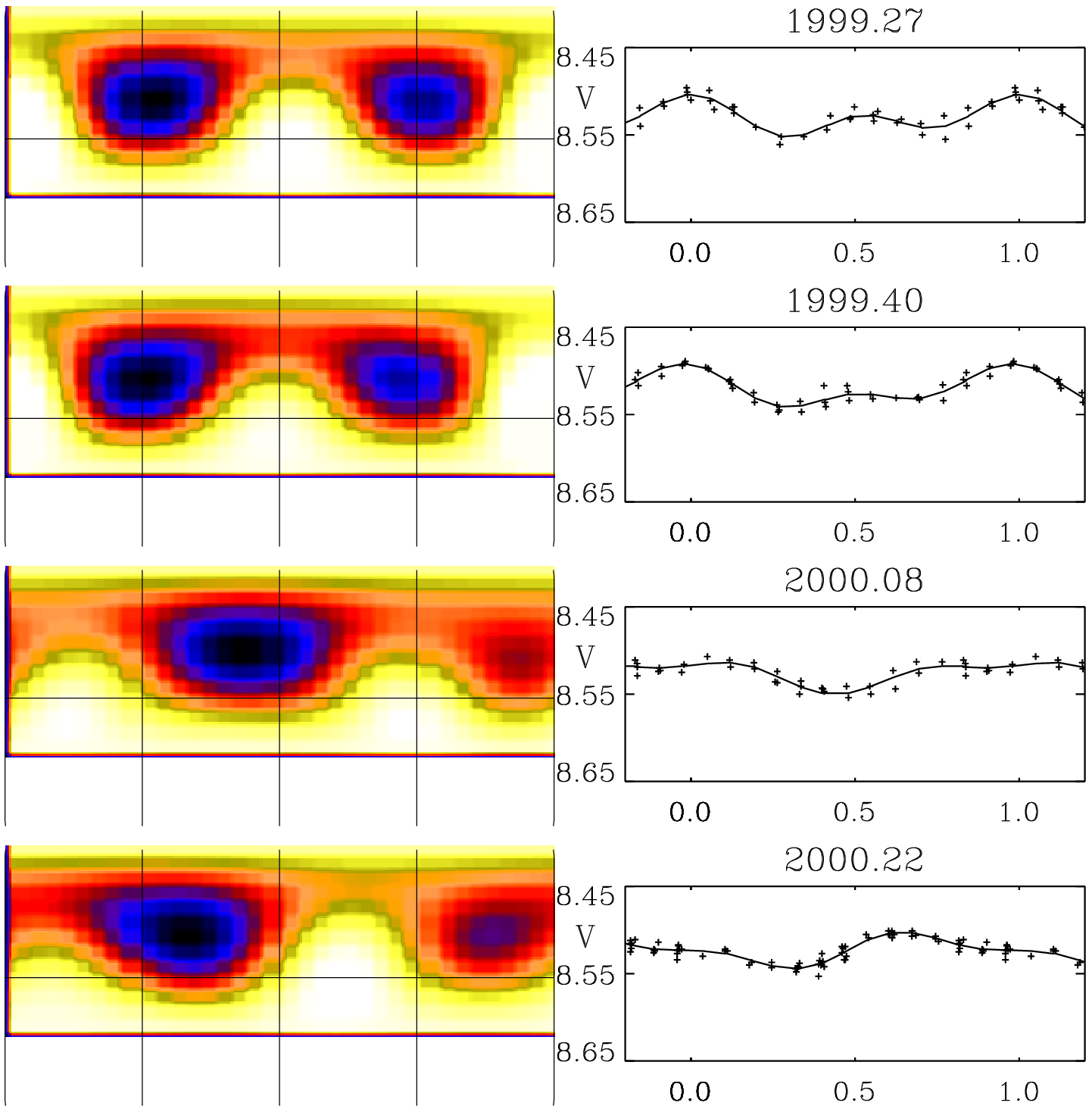} \includegraphics[width=4.5cm]{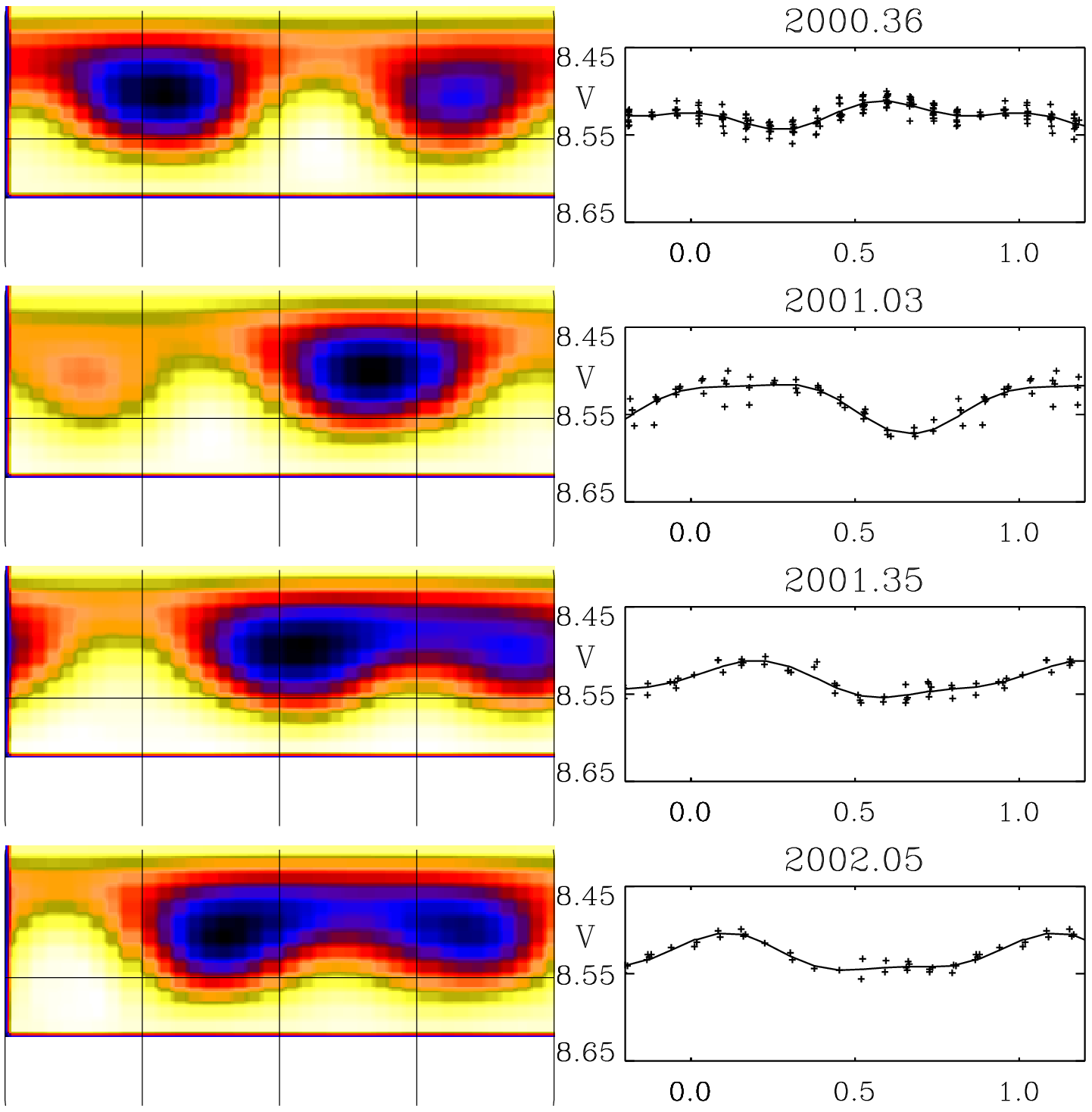} \includegraphics[width=4.5cm]{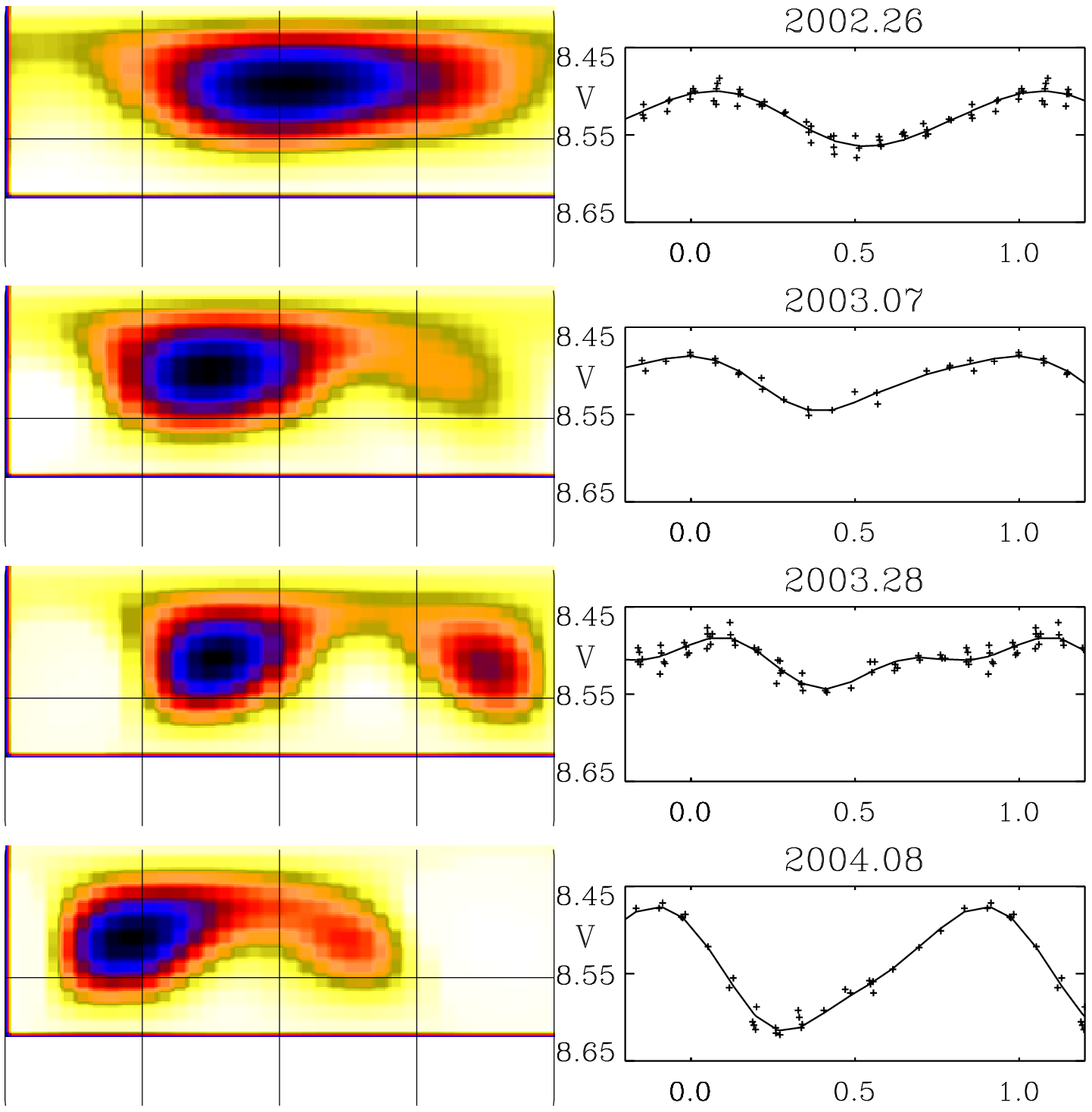} \includegraphics[width=4.5cm]{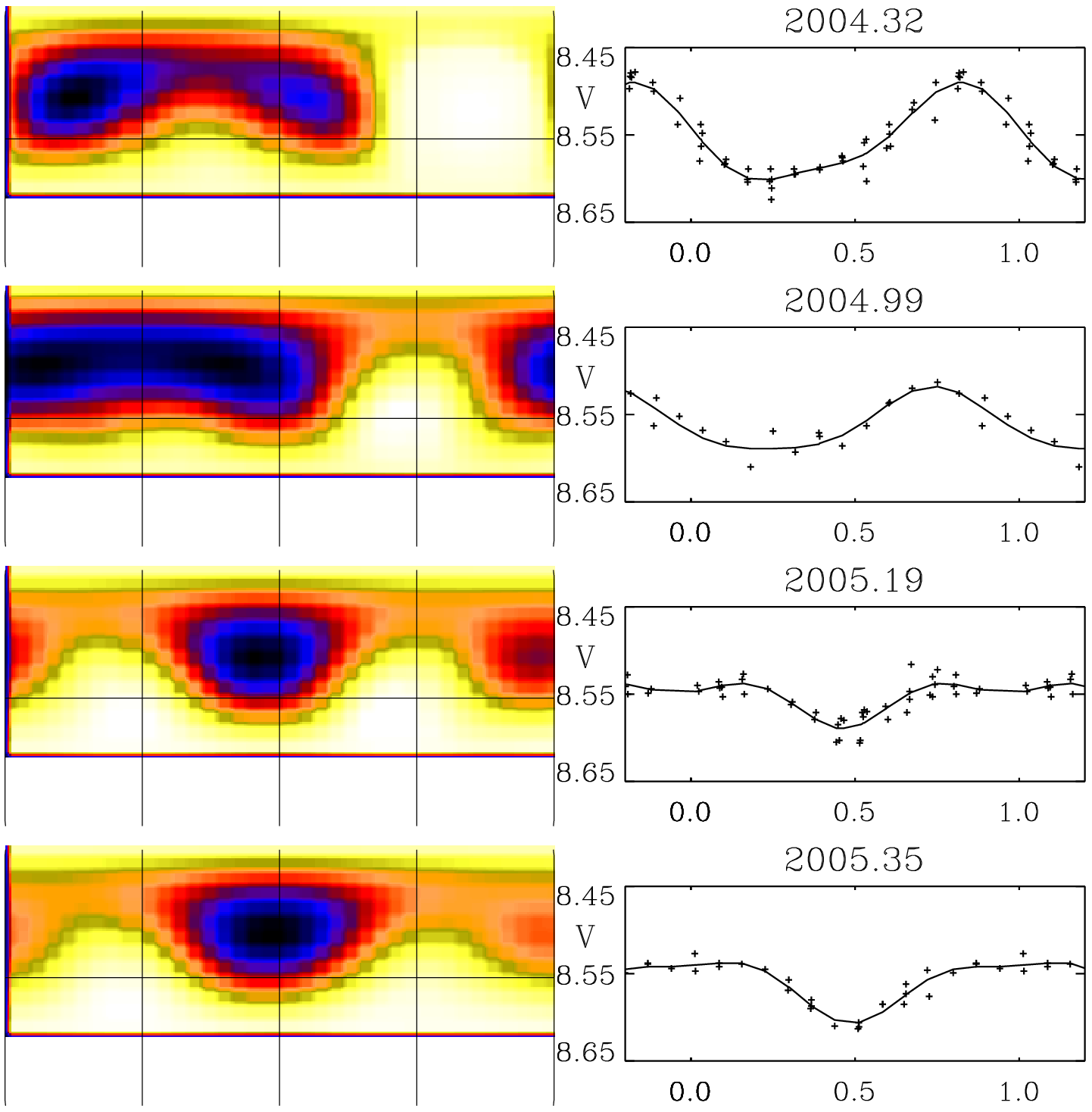}
\caption{Light curve inversions for \dpcvn. Darker regions indicate larger spot filling factors. The grid in the maps
indicates the equator and four longitudes separated by 90\degr. The observations (crosses) and calculated $V$ and $y$ light curves (solid lines) are also presented.
Note that for convenience reasons we use 2,451,250.99805 as the zero phase in the plot, i.e., the time of the first photometric observation
we have, yielding a phase shift of +0.240 as compared
to the Doppler maps in Sect.~\ref{sect_DI_dpcvn}).}
\label{F11}
\end{figure*}

The inversions of the individual light curves covering the interval 1999--2005 are presented in Fig.~\ref{F11} together with the light curves themselves.
As can be seen, the maps usually show two active regions which often are extended or even merged.   Also, these active regions seem to be quite stably separated by
half of the phase. This kind of activity pattern has been found on many active stars (see, e.g., Ol\'ah et al.~\cite{olah}, Henry et al.~\cite{henry}, Jetsu~\cite{jetsu2},
Berdyugina \& Tuominen \cite{ber_tuo}, etc.), and may bear also some evidence for the existence of a flip-flop phenomenon, where the
dominant part of the spot activity shifts 180$\degr$ in longitude over a short period of time and remains at this new active longitude
for some time, usually a few years (see e.g., Jetsu et al \cite{jetsu1}, Berdyugina \& Tuominen \cite{ber_tuo}, Korhonen et al.~\cite{kor}).

\subsection{Differential rotation}

Differential rotation can be assessed from seasonal photometric period changes, since the current period depends on the stellar latitude of the spots
that produce rotational modulation. Therefore, we derived rotation
periods for the ten available observing seasons (cf. Fig.~\ref{F2}). The values fall between $P_{\rm rot}^{\rm min}$=13.61\,d (2nd season) and
$P_{\rm rot}^{\rm max}$=14.39\,d (7th season). A rough guess of the surface shear parameter $\alpha$ can be estimated from $\Delta P/P$, getting $\approx$0,06.
We underline nonetheless,  that the sign of the surface shear (i.e., solar-type equatorial acceleration, or opposite) cannot be determined without additional information.
This result is quite close to the value derived from spectroscopy. On the other hand, the typical error of the period determination of 0.36\,d for such short intervals
makes this determination ambiguous.

To perform a more reliable analysis we divided the whole photometric dataset into two parts.
During the first half of the observing range (1st-5th seasons) the seasonal amplitudes were smaller, while the mean brightness level was higher,
as compared to the second half (6th-10th seasons) where the amplitudes were increased and the mean brightness was decreased.
This change can be interpreted by shifting dominant spotted region(s) towards lower latitudes. If so, we would expect longer rotation period
for the second half of the dataset when assuming antisolar-type differential rotation.
Fourier analysis resulted in $P_{\rm rot}^{\rm I}=13.862\pm0.035$\,d for the first term (1st-5th seasons),
and $P_{\rm rot}^{\rm II}=14.012\pm0.028$\,d for the second
term (6th-10th seasons), implying a minimum shear of $\alpha=-0.01$.
We conclude that such an overall period change is in agreement with the proposed antisolar-type differential rotation law for DP CVn in Sect.~\ref{sect_ccf}.

\section{Conclusions}\label{sect_concl}

The revised Hipparcos parallax places both \dpcvn\ and \dipsc\ at a position in the H-R diagram where a deep convective envelope coexists with a
hydrogen-burning shell. In that evolutionary phase, rapid redistribution of the moment of inertia must occur or had occurred (e.g. do Nascimento et al. \cite{donas})
and is likely to be a time of high dynamo activity due to the extra shear forces. Indeed, a number of overactive single giants were identified from their unusual high
Li abundance (e.g. Brown et al \cite{brown}, Fekel \& Balachandran \cite{fek:bal}). Therefore, these stars seem to have already experienced
their first Li dredge-up according to the models of Charbonnel \& Balachandran (\cite{cha:bal}).
Both stars of the present study likely have developed a degenerate helium core, and will later undergo a He-flash,
and  will evolve on to the early AGB and experience a second Li dredge-up.

The surface spot distribution in both stars is rather unspectacular in that there is no large polar cap, or a single
high-latitude feature or any other dominating feature. This is qualitatively in agreement with the absence of rotational
modulation in \Halpha\ at that time (for both stars, see Sect.~\ref{sect_halpha}), and suggestive of an atmosphere probably entirely penetrated and
saturated by magnetic fields. However, this may not be the case at all times, as shown by the light curve inversion for \dpcvn\  in 2002.26,
when spots were concentrated just in one hemisphere.

Both stars are very similar to the single high-Li K2-giant HD\,31993 (Strassmeier et al. \cite{hd31993}) in that they are very active and in
an evolutionary episode where we might expect a relation between its magnetic surface activity and the Li abundance. Knowing the
surface Li distribution and its relation to chromospheric activity indicators would be most useful for abundance determinations. Our
Doppler maps of HD\,31993 and the two stars in this paper may support the analogy with the Li enhancement in sunspots, i.e., that the
high Li abundance of these stars is related to the increased and nearly isotropic spottedness. Unfortunately, the quality of our spectra is only moderate and
higher S/N ratio and higher spectral resolution are warranted. Also, further Doppler images of Li-rich stars are needed to quantitatively elaborate on such a speculation.

\begin{acknowledgements}
Authors wish to thank an anonymous referee for his/her valuable comments.
We thank Thomas Hackman for analysing \dipsc\ photometry using TSPA period analysis method, Katalin Ol\'ah for the useful discussions on
{\sc MuFrAn} and Kriszti\'an Vida for his help in applying Lomb-method. ZsK and LK are grateful to the Hungarian Science
Research Program (OTKA) for support under the grant K-81421.
This work is supported by the ``Lend\"ulet-2009" Young Researchers' Program of
the Hungarian Academy of Sciences and by the
HUMAN MB08C 81013 grant of the MAG Zrt. HK acknowledges the support from the European Commission under the Marie Curie Intra-European Fellowship.
The authors acknowledge the support of the German \emph{Deut\-sche For\-schungs\-ge\-mein\-schaft, DFG\/} through projects KO~2320/1 and STR645/1.
KGS acknowledges the generous allotment of telescope time at KPNO as well as the access to the Canadian CFHT time. 
This study was supported by the Ministry of Education  of the Russian Federation through the Federal
Targeted program ``Scientific and Pedagogical Staff of Innovative Russia" for 2009–2013 (state contract 14.740.11.0084).
\end{acknowledgements}

%
%


\end{document}